\documentclass{aa}
\usepackage{natbib}
\bibpunct{(}{)}{;}{a}{}{,}% to follow the A&A style
\usepackage{graphicx}
\usepackage[figuresright]{rotating}
\usepackage{txfonts}
\usepackage{longtable}
\begin{document}
\newcommand{\water}{H$_2$O}
\newcommand{\solum}{\hbox{L$_{\sun}$}}
\newcommand{\wlum}{$L_{{\rm H_2 O}}$}
\newcommand{\kms}{km s$^{-1}$}
\newcommand{\HII}{\hbox{H{\sc ii}}}
\newcommand{\HI}{\hbox{H{\sc i}}}
\defcitealias{newmas}{HPT}
    \title{New \water\ masers in Seyfert and FIR bright galaxies. IV.}
          
   \subtitle{Interferometric follow-ups
\thanks {Table 5 is only available in electronic form at http://www.aanda.org}}
  % \subtitle{}

 \author{A.\ Tarchi\inst{1}
         \and P.\ Castangia\inst{1}
         \and C.\ Henkel\inst{2}
	 \and G. Surcis\inst{3,}\thanks{Member of the International Max Planck Research School (IMPRS) for Astronomy and Astrophysics at the Universities of Bonn and Cologne}
         \and K.\ M.\ Menten\inst{2}}

    \offprints{A.\ Tarchi}

 \institute{INAF-Osservatorio Astronomico di Cagliari, Loc. Poggio dei Pini, Strada 54, I-09012 Capoterra (CA), Italy\\
               \email{atarchi@ca.astro.it}
               \and Max-Planck-Insitut f\"ur Radioastronomie, Auf dem H\"ugel 69, 53121 Bonn, Germany
               \and Argelander-Institut f\"{u}r Astronomie der Universit\"{a}t Bonn, Auf dem H\"{u}gel 71, 53121 Bonn, Germany\\}
   
    \date{Received ; accepted }

% \abstract{}{}{}{}{}
% 5 {} token are mandatory

  \abstract
  {Very luminous extragalactic water masers, the megamasers, are associated with active galactic nuclei (AGN) in galaxies characterized by accretion disks, radio jets, and nuclear outflows. Weaker masers, the kilomasers, seem to be mostly related to star formation activity, although the possibility exists that some of these sources may belong to the weak tail of the AGN maser distribution.}
  {It is of particular importance to accurately locate the water maser emission to reveal its origin and shed light onto extragalactic star forming activity or to elucidate the highly obscured central regions of galaxies.}
  {We performed interferometric observations of three galaxies, NGC\,3556, Arp\,299, and NGC\,4151, where water emission was found. Statistical tools have been used to study the relation between OH and \water\ maser emission in galaxies.}
  {The maser in NGC\,3556 is associated with a compact radio continuum source that is most likely a supernova remnant or radio supernova. In Arp\,299, the luminous water maser has been decomposed in three main emitting regions associated with the nuclear regions of the two main galaxies of the system, NGC\,3690 and IC\,694, and the region of overlap. In NGC\,4151, only one of the two previously observed maser components has been tentatively detected. This feature, if real, is associated with the galaxy's central region. The only galaxy, so far, where luminous maser emission from two maser species, OH and \water\ has been confidently detected is Arp\,299. Weaker masers from these two species do instead coexist in a number of objects. A larger number of objects searched for both maser species are, however, necessary to better assess these last two results.}
  {}


   \maketitle
%
%________________________________________________________________
  
\section{Introduction}

While there is unanimous consent that the most luminous \water\ masers are related to the nuclear activity of their host galaxies (accretion-disks, nuclear jets, or outflows), the origins of the weaker extragalactic masers, i.\,e. the ``kilomasers'', are still a matter of debate.

Unlike the megamasers, the observed luminosities ({\wlum}\,$<$\,10\,{\solum}) of kilomasers can be explained by the superposition of several galactic W49N-like or even weaker sources. Furthermore, most of the known kilomasers are associated with particularly active star-forming regions clearly offset from the nucleus of their parent galaxy (e.\,g., M\,33, \citealt{m33}). This has led to the belief that kilomasers and megamasers may constitute two distinct classes of extragalactic {\water} masers. The former are in their majority related to star formation similarly to, although typically brighter than, the galactic masers, while the latter are thought to be exclusively powered by the nuclear activity of active galactic nuclei (AGN). The kilomaser sources associated with the active galactic nuclei of M\,51 \citep{m51} and of a few more {\water} masers of lower luminosity in the inner parsecs of NGC\,4051 (\citealt{n4051}), NGC\,520 (\citealt{castangia08}), and NGC\,3620 (\citealt{surcis09}) hint at the possibility that some kilomasers could also be related to nuclear activity, providing the low luminosity tail of the more powerful megamasers.

To date, beyond the Magellanic Clouds, the presence of 32 kilomasers has been reported from 24 galaxies (see Table~\ref{kilomasers}). The maser emission from 16 of these galaxies has not yet been studied at high resolution so that it is impossible to draw definite conclusions on its nature. This and the impact such measurements have on pinpointing spots of vigorous star formation or elucidating properties of nearby AGN has motivated, and still motivates, extended searches for more such masers as well as interferometric follow-ups to determine their origin.

With this goal in mind, we performed Very Large Array (VLA\footnote{The National Radio Astronomy Observatory (NRAO) is operated by Associated Universities, Inc., under a cooperative agreement with the National Science Foundation.}) observations of the kilomaser source NGC\,3556, detected in our far infrared (FIR) sample comprised of all galaxies with IRAS Point Source 100 $\rm \mu$m flux density $>$ 50 Jy and Dec. $>$ --30$^{\circ}$ (\citealt{newmas}, hereafter HPT). In addition, we used the VLA to observe the megamaser in Arp\,299, which is part of the same sample and which appears to arise from more than one spot (\citealt{tarchi07b}). Finally, we also present interferometric data from NGC\,4151, a Seyfert galaxy hosting a kilomaser with unusual features.
 
In Sects.\,2 and 3, we describe the observations and results, respectively. Sect.\,4, associates the detected masers with either star formation or AGN activity and analyses statistically the extragalactic \water\ and OH (kilo)masers known to date. Main conclusions are summarized in Sect.\,5. The present work is the last in a series of papers (\citealt{newmas}, \citealt{castangia08}, and \citealt{surcis09}) reporting searches for water maser emission in complete FIR-flux based samples of galaxies, highlighting new water maser line detections and presenting detailed maps of their surroundings.

%__________________________________________________________________

\section{Observations}\label{obs}

All galaxies were observed in Spectral Line Mode, in the $6_{16} \rightarrow 5_{23}$ transition of ortho-{\water} (rest frequency 22.23508\,GHz). 

\subsection{NGC\,3556}
NGC\,3556 was observed with the VLA in two hybrid configurations, CnB and DnA, on September 28, 2002 and January 25, 2006, respectively. In both cases, observations were made employing a single band of width 6.25\,MHz centered at the velocity of the maser feature detected with Effelsberg (740\,{\kms}; HPT). The observing band was subdivided into 128 channels each of width 48.8\,kHz, which corresponds to $\sim$\,0.7\,{\kms}. The flux density scale was determined in both observing runs by measuring the non-variable source 3C\,286, with an adopted flux density of 2.54\,Jy, as calculated by the NRAO's Astronomical Image Processing System (AIPS) task SETJY using the revised (by R.\ Perley) scale of \cite{baars77}. The phase calibrator was 11282+59252, whose flux density was estimated to be 0.56$\pm$0.01\,Jy and 0.65$\pm$0.01\,Jy in the two runs, respectively. We used 3C\,286 also to determine the bandpass corrections. On January 5, 2008, NGC\,3556 was also observed at 1.4 GHz by the Multi-Element Radio Linked Interferometer Network (MERLIN\footnote{MERLIN is a National Facility operated by the University of Manchester at Jodrell Bank Observatory on behalf of STFC}) in Wide-field Continuum Mode for  a total on-source integration time of 18 hours. The sources 3C\,286 and 1055+567 were used as flux and phase calibrators, respectively.

\subsection{Arp\,299}
Water vapor in Arp~299 was observed on September 19, 2004, and April 9, 2007, with the VLA in its A and D configurations, respectively. A frequency setup with two 25 MHz intermediate frequency bands (``IFs'') centered at Local Standard of Rest velocities of $V_{\rm LSR}$ = 2900 and 3200 \kms was used. The IFs were overlapped in frequency in order to minimize the effect of band-edge roll-off. Each IF covered $\sim$ 340 $\rm km\:s^{-1}$ with a resolution of $\sim$ 20 $\rm km\:s^{-1}$. For both datasets, the flux density and bandpass calibration was performed by using 3C\,286 (2.56\,Jy) and the phase calibration was derived from observations of the point source 11282+59252 (0.53$\pm$0.01 and 0.30$\pm$0.01\,Jy for the A and D array, respectively). Radio continuum maps were produced using the (few) line-free channels. For the A-array data, the restoring beam was 0\farcs1 $\times$ 0\farcs1 and the rms noise was 0.5 mJy/beam/chan and 0.25 mJy/beam for the cube and the continuum maps, respectively. The noise was higher than the expected thermal noise, likely because of poor weather during the observations. For the D-array data, the restoring beam was 4\farcs9 $\times$ 4\farcs0 and the rms noise was 0.25 mJy/beam/chan and 0.1 mJy/beam for individual spectral channels and the continuum maps, respectively. Spectral line 22-GHz single-dish observations of the water maser in Arp\,299 were also performed with the 100-m Effelsberg radio telescope in November 2005 by pointing at the three main centers of activity in the system (see Sect.\,3.2) and in April 5, 2007 (only four days before the VLA D-array measurements) with a single pointing intermediate between the three aforementioned locations.

\subsection{NGC\,4151}
NGC\,4151 was observed with the VLA in its DnA hybrid configuration on January 25, 2006. Observations were made using two 3.125-MHz IFs centered at the velocity of the two maser features detected with the GBT (692 and 1127 \,{\kms}, respectively; \citealt{braatz04}). Each IF was subdivided into 128 channels that provide a channel spacing of 24.4\,kHz corresponding to $\sim$\,0.33\,{\kms}. The flux density scale and bandpass corrections were determined by using 3C\,286 (2.54\,Jy). The phase calibration was derived from observations of 11470+39586 (0.88$\pm$0.01\,Jy).

\subsection{Data reduction and position accuracy}\label{sect:posit}
The data reduction was made using AIPS. All datasets were calibrated in the standard way. The radio continuum emission was subtracted from the spectral line data using AIPS task UVLSF, that fits a straight line to the visibilities of the line-free channels and then subtracts it from the \textit{uv} dataset. This task also provides the fitted baseline as an \textit{uv} dataset that was used to create continuum maps. Each individual dataset was Fourier-transformed using natural weighting and then deconvolved utilizing the CLEAN algorithm (\citealt{hoegbom74}). Details of the observations and of the interferometric maps are summarized in Table \ref{vla_obs}.

The accuracy on absolute positions in a map due to statistical errors can be estimated using the synthesized beam size divided by the signal-to-noise ratios (for details on this derivation see, e.g., \citealt{hagi01}). For our observations taken in the VLA D and CnB-array configurations, due to the relatively coarse resolution, the statistical errors dominate, and hence, they are representative of the positional accuracy for the maps produced. For the VLA A and DnA-array maps, the positional uncertainties are dominated by the error due to the different position of the phase calibrator w.r.t. that of the target galaxies. The nominally estimated error for absolute positions in this case is of 0\,\farcs1. Since in our measurements we have used the "fast switching" technique and since errors can be divided by the signal-to-noise ratio (see above), this value can be taken as a safe upper limit (see, e.g., \citealt{tarchi07a}). Coordinates of the maser peaks in our VLA maps and associated positional errors are reported in Table~\ref{vla_lines}.

The relative positions between line and continuum peak positions derived from the same data is only limited by the signal-to-noise ratios since the uncertainty due to the different position of the calibrator w.r.t. that of the target galaxy affects both positions in the same way and cancels out. Hence, for our data we estimate the error on relative positions to be $\sigma_{\rm rel} = \sqrt{(\theta_{l}/(2 \cdot SNR_{l}))^{2} + (\theta_{c}/(2 \cdot SNR_{c}))^{2}}$, where $\theta$ denotes the restored beam size of the map in use, and l and c refer to line and continuum emission, respectively (for details, see also \citealt{tarchi07a}, and references therein). When the line and peak positions are instead derived by different datasets, the final accuracy can be estimated by the the quadratic sum of the errors on the absolute positions in the two maps.

\begin{table*}
%{\scriptsize{
%\centering
\caption{Interferometric maps.}
\label{vla_obs}
\vspace{0.2cm}
\begin{minipage}[t]{\textwidth}
\renewcommand{\footnoterule}{}
%\hspace{-0.5cm}
\begin{tabular}{lcccccccccc}
\hline\hline

Source    & Epoch      & Telescope & Mode & Freq. & Bandwidth & IFs & Channel width & Weighting & Beam size &       r.m.s.       \\
          &            &           &      & (GHz)     &   (MHz)   & No. &    (kHz)      &           &           & (mJy\,beam$^{-1}$chan$^{-1}$) \\

\hline 

NGC\,3556 & 28-Sep.-02 &  VLA-CnB  & Line &   22    & 6.25          & 1 &  48.8          & Natural   & 1\farcs15$\times$0\farcs43 & 4   \\
NGC\,3556 & 25-Jan.-06 &  VLA-DnA  & Line &   22        & 6.25      & 1 &  48.8          & Natural   & 0\farcs27$\times$0\farcs19 & 3.5   \\
NGC\,3556 & 5-Jan.-08 &  MERLIN  & Cont. &   1.4        & 16      & 1 &  --          & Natural   & 0\farcs22$\times$0\farcs17 & 0.04   \\
NGC\,3556 & 5-Jan.-08 &  MERLIN  & Cont. &   1.4        & 16      & 1 &  --          & Uniform   & 0\farcs16$\times$0\farcs14 & 0.05   \\
Arp\,299  & 19-Sep.-04 &  VLA-A   & Line &    22         & 16.0      & 2 & 1562.5         & Natural   & 0\farcs10$\times$0\farcs10 & 0.5 \\
Arp\,299  & 9-Apr.-07 &  VLA-D   & Line &    22         & 16.0      & 2 & 1562.5         & Natural   & 4\farcs89$\times$4\farcs03 & 0.25 \\
NGC\,4151 & 25-Jan.-06 &  VLA-DnA & Line &    22         & 3.125     & 2 &  24.4          & Natural   & 0\farcs24$\times$0\farcs11 & 10   \\
\hline
\end{tabular}
\end{minipage}
%}}
\end{table*}

%
%______________________________________________________________

\section{Results}\label{res}
  
\subsection{NGC\,3556}\label{sect:n3556}

\begin{figure}
\centering
\includegraphics[width= 8.5 cm]{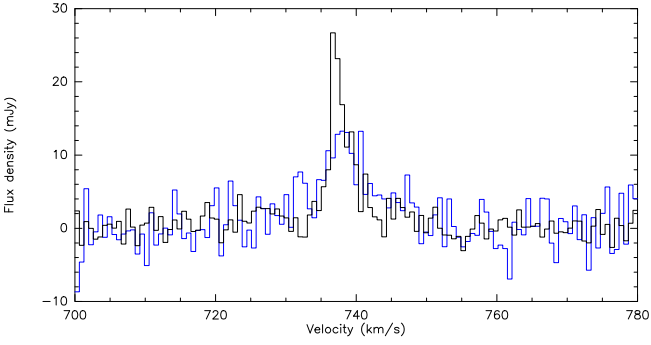}
\caption{The VLA CnB-array (blue solid line) and DnA-array (black solid line) water maser spectra of NGC\,3556 taken in 2002 and 2006, respectively. Both spectra have a channel spacing of 0.7 \kms.}
\label{fig:n3556_duespec}
\end{figure}

\begin{figure}
\centering
\includegraphics[width= 8 cm]{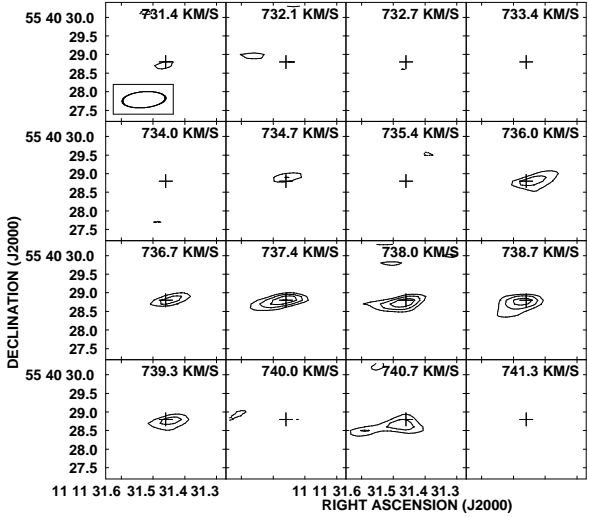}
\caption{Channel maps of the {\water} maser emission in NGC\,3556 observed with the VLA CnB array. Contour levels are -3, 3, 3.5, 4\,$\times$4\,mJy\,beam$^{-1}$ (1$\sigma$ rms = 4\,mJy\,beam$^{-1}$). The cross indicates the position of the line peak.}
\label{fig:kntr_2}
\end{figure}

\begin{figure}
\centering
\includegraphics[width= 8 cm]{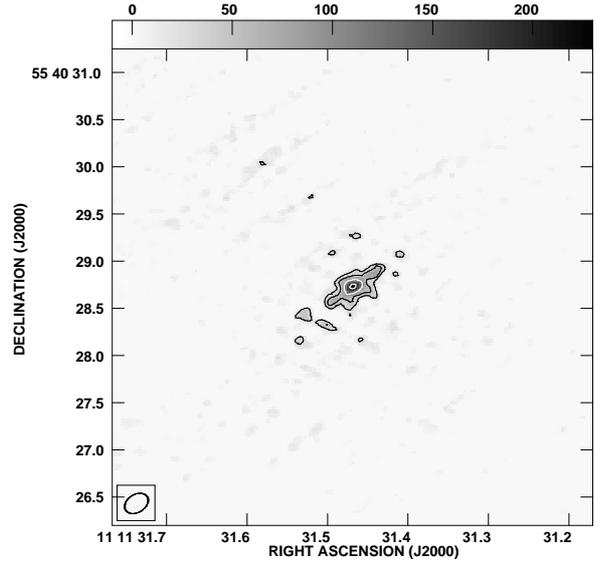}
\caption{Grey scale and contour plot of the total integrated intensity of the maser emission in NGC\,3556 observed with the VLA DnA array. Integration was performed over the velocity interval 720--760\,{\kms}. Contours: -1,1,2,4 $\times$ 25 mJy/beam/\kms. (1$\sigma$ rms = 0.4 mJy/beam/\kms).}
\label{fig:mom0_2}
\end{figure}

%\begin{figure}
%\centering
%\includegraphics[width= 8.5 cm]{NGC3556/kntr_cnb.ps}
%\caption{Channel maps of the {\water} maser emission in NGC\,3556 observed with the VLA CnB array.Contour levels are -3, 3, 3.5, 4\,$\times$4\,mJy\,beam$^{-1}$. The cross indicates the position of the line peak.}
%\label{fig:n3556_chmap_cnb}
%\end{figure}

%\begin{figure}
%\centering
%\includegraphics[width= 8.5 cm]{NGC3556/kntr_dna.ps}
%\caption{Channel maps of the {\water} maser emission in NGC\,3556 observed with the VLA DnA array. Contour levels are -3, 3, 4, 5, 6, 7, 8\,$\times$4\,mJy\,beam$^{-1}$. The cross indicates the position of the line peak.}
%\label{fig:n3556_chmap_dna}
%\end{figure}

%\begin{figure}
%\centering
%\includegraphics[width= 8.5 cm]{NGC3556/mom0_cnb.ps}
%\caption{Grey scale and contour plot of the total integrated intensity of the maser emission in NGC\,3556 observed with the VLA CnB array. Integration was performed over the velocity interval 720--760\,{\kms}. Contours: -1,1,2,4 $\times$ 30 mJy/beam/\kms.}
%\label{fig:n3556_mom0_cnb}
%\end{figure}

%\begin{figure}[htpb]
%\centering
%\includegraphics[width= 8.5 cm]{NGC3556/mom0_dna.ps}
%\caption{Grey scale and contour plot of the total integrated intensity of the maser emission in NGC\,3556 observed with the VLA DnA array. Integration was performed over the velocity interval 720--760\,{\kms}. Contours: -1,1,2,4 $\times$ 25 mJy/beam/\kms.}
%\label{fig:n3556_mom0_dna}
%\end{figure}

NGC\,3556 is a nearby\footnote{The distance is taken from \citet{n3556king97} and is derived from the radial velocity corrected for the anisotropy of the 3K cosmic microwave background.} (D\,$\sim$\,12\,Mpc, 1\arcsec is equivalent to $\sim$\,60\,pc) 
edge-on spiral galaxy with a FIR luminosity similar to that of the Milky Way ($L_{\rm FIR} \sim 10^{10}$\,{\solum}). 
The presence of a conspicuous radio halo \citep{n3556irwin99}, prominent {\HI} loops interpreted as expanding supershells 
\citep{n3556king97}, and extraplanar diffuse X-ray emission \citep{n3556wang03} all indicate that NGC\,3556 is undergoing 
an intense disk-halo interaction. 
At high resolution, the radio continuum emission is resolved into a number of discrete components, which form a partial ring around the 
{\HI} kinematic center (\citealt{n3556irwin00}; hereafter ISE). 
Although a radio core is absent, Chandra X-ray observation revealed an ultraluminous X-ray source close to the {\HI} kinematic center 
and the 2${\rm \mu\,m}$ peak, which has a power law spectrum typical of an AGN \citep{n3556wang03}. 
A significant amount of dense molecular gas is present in the galaxy, as proved by $^{12}$CO and HCN observations \citep{n3556gao04}. 
Water maser emission has been detected, for the first time, with Effelsberg, in March 2002. The detection spectrum and three more single-dish spectra at subsequent epochs are shown in HPT (their Fig.\,5).

In our VLA observations, we detect {\water} water maser emission at $V_{\rm LSR} \sim 738$\,{\kms} with a peak flux density for the CnB and DnA arrays of $\sim$\,13 and 27 mJy, respectively (Fig.~\ref{fig:n3556_duespec}). A Gaussian fit to the spectra indicates that the full width to half maximum (FWHM) of the maser feature is 8$\pm$2 and 3.7$\pm$0.3 \kms, yielding for the two arrays similar values for the isotropic maser luminosity, 0.29 and 0.28\,{\solum} (see Table~\ref{vla_lines}).

\begin{table*}
%{\scriptsize{
\centering
\caption{Parameters of the maser lines observed with the VLA.}
\label{vla_lines}
%\vspace{0.2cm}
\begin{minipage}[t]{\textwidth}
\renewcommand{\footnoterule}{}
%\hspace{-0.5cm}
\begin{tabular}{lcccccccc}
\hline\hline

Source  & Distance  & Array & \multicolumn{2}{c}{Peak coordinates\footnote{Error estimates associated with this quantity are discussed in Sect.~\ref{sect:posit}}} & $V_{\rm LSR,opt}$\footnote{All values have been obtained by Gaussian fits to the maser line(s) with the exception of NGC\,4151 where a Gaussian fit would have been meaningless. In this latter case, the channel width of the spectrum has been taken as a reliable upper limit for the uncertainty in the peak velocity, and the error on the line integrated intensity has been computed assuming errors of 10 and 20 \% in the line intensity and width, respectively.} & $\int{S {\rm d}V}^{b)}$ & {\wlum}\footnote{{\wlum}/[{\solum}] = 0.023 $\times$ $\int{S {\rm d}V}$/[Jy\,{\kms}] $\times$ $D^2/[\rm Mpc^2]$.} & Notes\\
        &   (Mpc)  & &  RA & Dec.  & ({\kms}) & (Jy\,{\kms}) & ({\solum}) & \\  

\hline 

NGC\,3556         & 12 & CnB &    11 11 31.46\,$\pm$\,0.02 & +55 40 28.8\,$\pm$\,0.2 & 738.2\,$\pm$\,0.7 &  0.087\,$\pm$\,0.004  &  0.29\,$\pm$\,0.01  &  \\
NGC\,3556         & 12 & DnA &    11 11 31.47\,$\pm$\,0.01 & +55 40 28.7\,$\pm$\,0.1 & 737.5\,$\pm$\,0.1 &  0.083\,$\pm$\,0.005  &  0.28\,$\pm$\,0.01 &  \\
Arp\,299          &    &     &                 &                           &       &    &              &  \\
-- IC\,694         & 42 & A   &   11 28 33.65\,$\pm$\,0.01 & +58 33 46.8\,$\pm$\,0.1 & 2996\,$\pm$\,8   &  0.16\,$\pm$\,0.03  &  6.4\,$\pm$\,1.2  &    \\
                  &    & D   &    11 28 33.65\,$\pm$\,0.01 & +58 33 46.8\,$\pm$\,0.1 & 2988\,$\pm$\,2   &  0.46\,$\pm$\,0.03  &  19\,$\pm$\,1 &  \\
-- NGC\,3690       & `` & A   &   11 28 30.99\,$\pm$\,0.01 & +58 33 40.7\,$\pm$\,0.1 & 3112\,$\pm$\,8   &  0.41\,$\pm$\,0.05 &  17\,$\pm$\,2 &   \\
                  &    & D   &    11 28 30.97\,$\pm$\,0.02 & +58 33 40.8\,$\pm$\,0.2 & 3112\,$\pm$\,4   &  0.55\,$\pm$\,0.03 &  22\,$\pm$\,1 & \\
-- C+C'-complex       & `` & A   &   11 28 30.63\,$\pm$\,0.01 & +58 33 52.4\,$\pm$\,0.1 & 3179\,$\pm$\,18  &  0.18\,$\pm$\,0.04  &   7.5\,$\pm$\,1.5 & tentative   \\
$\:$$\:$$\:$C'     &    & D  &    11 28 31.35\,$\pm$\,0.02 & +58 33 49.8\,$\pm$\,0.2 & 3158\,$\pm$\,4   &  0.30\,$\pm$\,0.04  &  12\,$\pm$\,2 &    \\
$\:$$\:$$\:$C     &    & D   &    11 28 30.65\,$\pm$\,0.11 & +58 33 49.3\,$\pm$\,0.9 & 3142\,$\pm$\,7   &  0.12\,$\pm$\,0.03  &    4.9\,$\pm$\,1.0 & tentative\\
NGC\,4151         & 13 & DnA &    12 10 32.58\,$\pm$\,0.01 & +39 24 21.1\,$\pm$\,0.1 & 695.6\,$\pm$\,0.3 & 0.009\,$\pm$\,0.002 & 0.035\,$\pm$\,0.007 & tentative\\
\hline
\end{tabular}
\end{minipage}
%}}
\end{table*}

Comparing the VLA CnB spectrum observed on September 28, 2002,
with the spectrum taken with Effelsberg one day before (HPT; upper panel of their Fig.\,5),
we note that the peak flux density of the maser line observed by the VLA is less than half of that observed with Effelsberg. The intensity of the VLA DnA spectrum is, instead, closer to that of the single-dish measurements.  
For all observations, taken with either Effelsberg or the VLA, the $V_{\rm LSR}$ velocity and linewidth of the maser features are consistent with each other. 

Figures \ref{fig:kntr_2} and \ref{fig:mom0_2} show contour plots of those channels, which exhibit significant line emission, and a moment-0 map (i.e. a map of the velocity-integrated intensity) in the velocity range 720--760\,{\kms} obtained from the two CnB and DnA VLA datasets, respectively.
These maps indicate that the maser emission is slightly resolved both in space and velocity.

We fitted a two dimensional Gaussian to the brightness distribution of the source in the VLA CnB channel map showing the most 
intense emission (from 737.4 to 738.7 \kms in Fig. \ref{fig:kntr_2}, left panel) 
and find that the bulk of the emission arises from a spot of beam deconvolved dimensions of $\sim$\,0\farcs7\,$\times$\,0\farcs2 at position
$\alpha_{\rm J2000}$=11$^{\rm h}$11$^{\rm m}$31\fs46 and $\delta_{\rm J2000}$=55\degr40\arcmin28\farcs8.
Tentative maser emission also seems to be present in a spot $\sim$\,70 pc east of the main feature at a velocity $V_{\rm LSR}=740.7$\,{\kms} and position $\alpha_{\rm J2000}$=11$^{\rm h}$11$^{\rm m}$31\fs60 and $\delta_{\rm J2000}$=55\degr40\arcmin28\farcs5 (Fig.~\ref{fig:kntr_2}).
%and the latter at $V_{\rm LSR}=748.6$\,{\kms} and with coordinates $\alpha_{\rm J2000}$=11$^{\rm h}$11$^{\rm m}$31\fs35 and $\delta_{\rm J2000}$=55\degr40\arcmin28\farcs0. 
Additional weak spots might be present both in the CnB and DnA channel maps. These, however, are faint, arise from isolated channels, and may be spurious.

Since VLA broad band 22\,GHz data were not available, we produced CnB and DnA-array VLA continuum maps from the line-free channels. In both maps, no continuum emission was detected above a 1\,$\sigma$ rms noise of $\sim$ 0.5\,mJy\,beam$^{-1}$. This result is consistent with the 1.4-GHz flux densities reported by ISE for the compact sources in the nuclear region of NGC\,3556. Even the strongest source(s) of ISE remain(s) undetected in our maps indicating a non-thermal origin of the emission.

\subsection{Arp\,299}\label{sect:a299}

The extremely luminous infrared galaxy Arp~299 (Mkn~171)\footnote{According to the NASA/IPAC Extragalactic Database (NED), the entire merger forms NGC 3690, while IC 694 is a less prominent galaxy 1$^{\prime}$ to the northwest. Here we follow the more traditional nomenclature that is commonly used in the literature.} is a merging system located at a distance of $\sim$42\,Mpc \citep{casoli99}. It is composed of three main regions of activity: the two galaxies IC~694 ({\bf A}) and NGC~3690 ({\bf B}; resolved at infrared and radio wavelengths into two components, {\bf B1} and {\bf B2})\footnote{IR, radio, and X-ray observations all indicate that the putative nucleus of NGC\,3690 is located at position B1. Source B2 is more prominent in the optical because it is less obscured than B1 at visible wavelengths and may be associated with a region vigorously forming massive stars (\citealt{alonso09}, and references therein). Hereafter, we will therefore consider B1 as the nucleus of NGC\,3690.}, and the individual concentrations ({\bf C} and {\bf C$^{\prime}$}) at the interface where IC~694 and NGC~3690 overlap (for a description of the system and for the nomenclature used, see, e.g., \citealt{neff04}; their Fig.~2). The system is rich in molecular gas \citep{casoli99} and displays OH megamaser activity with an isotropic luminosity of $\sim$ 240 \solum\ \citep{baan85} apparently tracing a rotating disk in IC~694 (\citealt{bh90}; hereafter BH90). The presence of such a flattened rotating structure in IC~694 was subsequently also invoked by \citet{polatidis01} to explain its atomic (\HI) and molecular (CO) gas velocity distribution. In March 2002, strong water megamaser emission was detected in the merging system Arp 299, with a total isotropic luminosity of $\sim$ 200 L$_\odot$ (HPT). Later, preliminary results of our VLA A-array observations were reported in \citet{tarchi07b}. They show that the water maser emission in the system is produced mainly by two spots associated with the innermost regions of IC~694 and NGC~3690. Tentative emission was seen in the overlapping region. 

\begin{figure*}
\includegraphics[width= 18.2 cm]{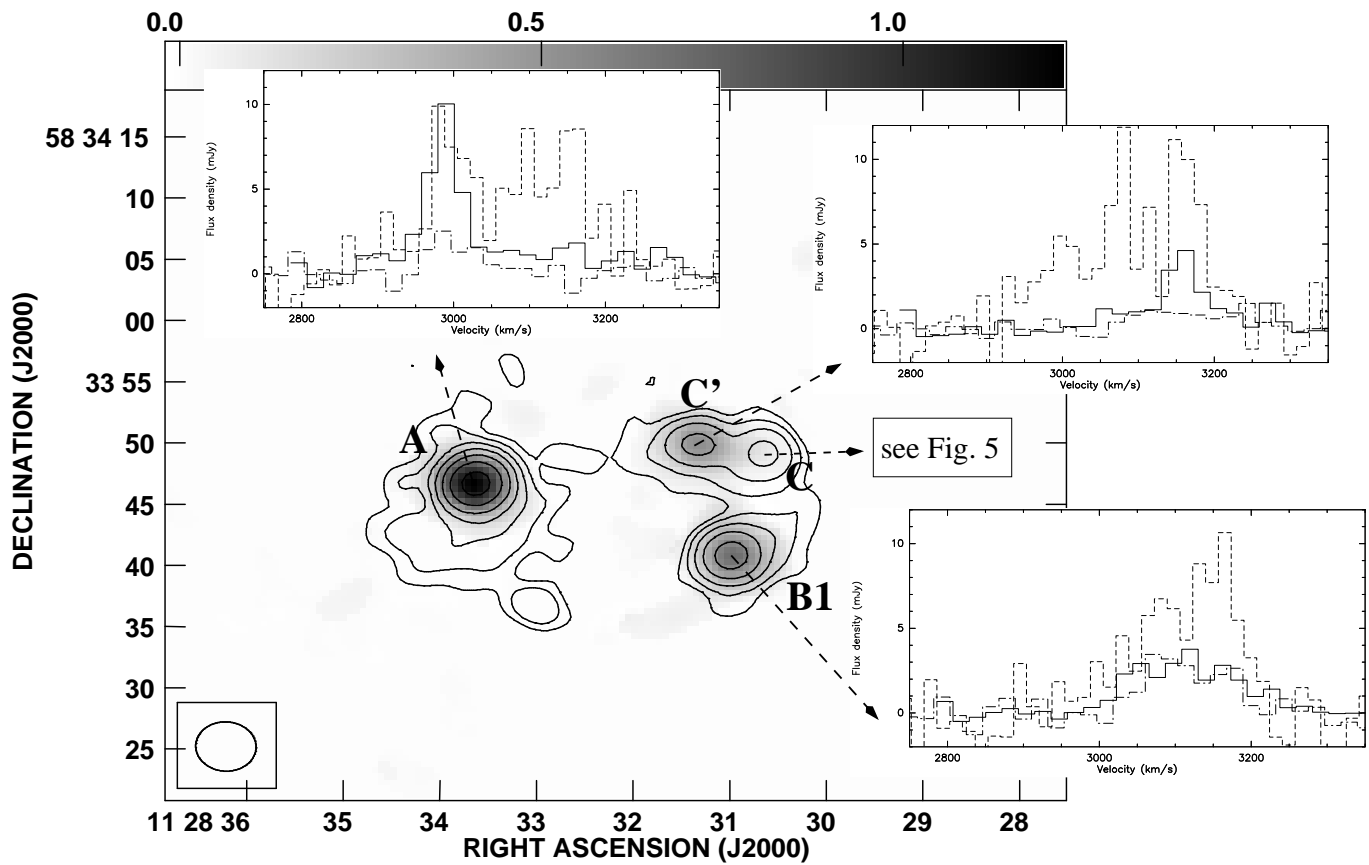}
\caption{
\textit{Big panel}: moment-zero map (grey scale) of the water maser emission in the Arp299 system observed by the VLA in its D configuration superimposed on the 22-GHz uniformly-weighted VLA D-array radio continuum image (contours) of the system produced by using the line-free channels in the maser data cube. Contour levels are -1, 1, 2, 4, 8, 16, 32, 64\,$\times$0.5\,mJy\,beam$^{-1}$ (1$\sigma$ rms = 0.15\,mJy\,beam$^{-1}$). \textit{Small panels}: for each center of maser emission within the Arp299 system, spectra from Effelsberg (dashed line), VLA D-array (solid line), and VLA A-array (dash-dotted line) are shown, with a channel spacing of 17, 20, and 20 \kms, respectively.} 
\label{fig:arp299_tot}
\end{figure*}

\begin{figure}
\includegraphics[width= 8.5 cm]{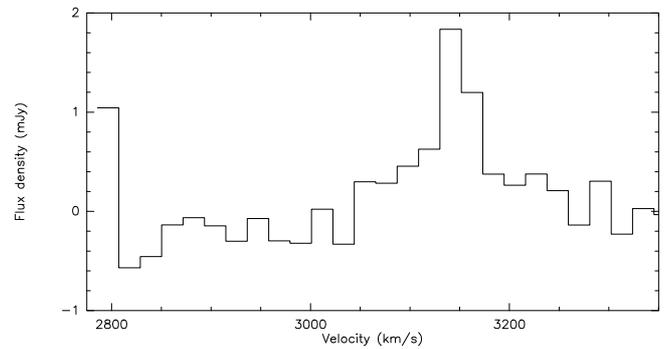}
\caption{Weak water maser feature detected in the Arp299 system observed by the VLA in its D configuration at the position of the western peak of the C+C'-complex, labeled C (\citealt{neff04}; Fig.~\ref{fig:arp299_tot}). The channel spacing is 20 \kms.} 
\label{fig:over_bis}
\end{figure}

In a more thorough data reduction of the VLA A-array data, the scenario described in \citet{tarchi07b} is confirmed with higher confidence (see Fig.~\ref{fig:arp299_tot}). Luminous maser emission is detected at two locations with coordinates $\alpha_{\rm J2000}$=11$^{\rm h}$28$^{\rm m}$33\fs65; $\delta_{\rm J2000}$=58\degr33\arcmin46\farcs8 and $\alpha_{\rm J2000}$=11$^{\rm h}$28$^{\rm m}$30\fs99; $\delta_{\rm J2000}$=58\degr33\arcmin40\farcs7, corresponding to the two brightest centers of activity in the system labeled A and B1 by \citet{neff04}. The maser features have peak velocities of 2996 and 3112 \kms, respectively. Our recent VLA D-array data confirm the presence of emission in locations and velocities consistent, within the accuracies, with those resulting from the previously taken higher resolution VLA data. In particular, the tentative emission detected with the A-array in the overlap region is convincingly confirmed. Most of the emission originates from a bright spot ($\sim$ 4 mJy), peaks at a velocity of 3158 \kms, and has a position, $\alpha_{\rm J2000}$=11$^{\rm h}$28$^{\rm m}$31\fs35; $\delta_{\rm J2000}$=58\degr33\arcmin49\farcs8, coincident with the source C' of \citet{neff04}. In Fig.~\ref{fig:arp299_tot} (large panel), we show the moment-zero map of the water maser emission in Arp299 observed by the VLA in its D configuration superposed on the 22-GHz uniformly-weighted VLA D-array radio continuum image produced by using the line-free channels in the maser data cube. In the smaller panels of the same figure, the spectra of the maser emitting regions observed with Effelsberg and the VLA, in its A and D configurations, are plotted. Location, line peak velocity, flux density, linewidth, and isotropic luminosity of the maser features are given in Table~\ref{vla_lines}. Weaker emission ($\sim$ 1.8 mJy) in the overlap region is also detected at a position of $\alpha_{\rm J2000}$=11$^{\rm h}$28$^{\rm m}$30\fs65; $\delta_{\rm J2000}$=58\degr33\arcmin49\farcs3 with a peak velocity of 3142 \kms\ and a line width of $\sim$ 25 \kms. The location is coincident with the source C of \citet{neff04}. For the sake of clarity, this feature is not included as a panel in Fig.~\ref{fig:arp299_tot} while it is shown, using a narrower, more-suitable y-axis scale, in Fig~\ref{fig:over_bis}. 

\subsection{NGC\,4151}\label{sect:n4151}

NGC~4151 is a grand-design, weakly barred spiral located at 13.3 Mpc \citep{braatz04}, which is known to host an active galactic nucleus (AGN). Although it is often classified as a type 1.5 Seyfert \citep{osterbrock76}, its broad optical emission lines are variable and sometimes give a type 2 Seyfert spectrum (\citealt{sergeev01}). The equivalent hydrogen column density, inferred from the X-ray spectrum toward the nucleus, is also similar to that found in typical Seyfert 2 galaxies ($N_{\rm{H}} \sim 1.5 \times 10^{23}$~cm$^{-2}$, Wang et al. 2010). 
Radio observations of the nucleus show a linear radio structure (jet) approximately 3$\,.\!\!^{\prime\prime}$5 (220~pc) in length, at a position angle (P.A.) 77$^{\circ}$ \citep{wilson82}. \citet{carral90} identified six discrete knots along the jet at 15~GHz, the strongest of which (peak flux density: 12.5~mJy/beam) has a flat spectral index which is consistent with synchrotron self absorption associated with an AGN. Higher angular resolution VLBA and phased VLA observations at 1.4~GHz resolved the strongest knot into several components that appear to be associated with changes in the orientation of a faint highly collimated jet (diameter $\leq$1.4~pc), as it interacts with small gas clouds (\citealt{mundell03}). The identification of one of these components with the AGN is still uncertain. In \citet{braatz04}, the detection of water maser emission with an isotropic luminosity of 0.7 \solum\ was reported. The maser emission is confined primarily to two very narrow components, one at 692.4 \kms\ and the other at 1126.6 \kms, blushifted and redshifted, respectively, w.r.t. the systemic velocity of the galaxy, 995 \kms. For a channel spacing of 0.33 \kms, Gaussian fits of the emission lines yielded peak flux densities and linewidths of 36 and 53 mJy, and 1.2 and 1.5 \kms (FWHM), respectively. 

\begin{figure*}
\includegraphics[width= 17 cm]{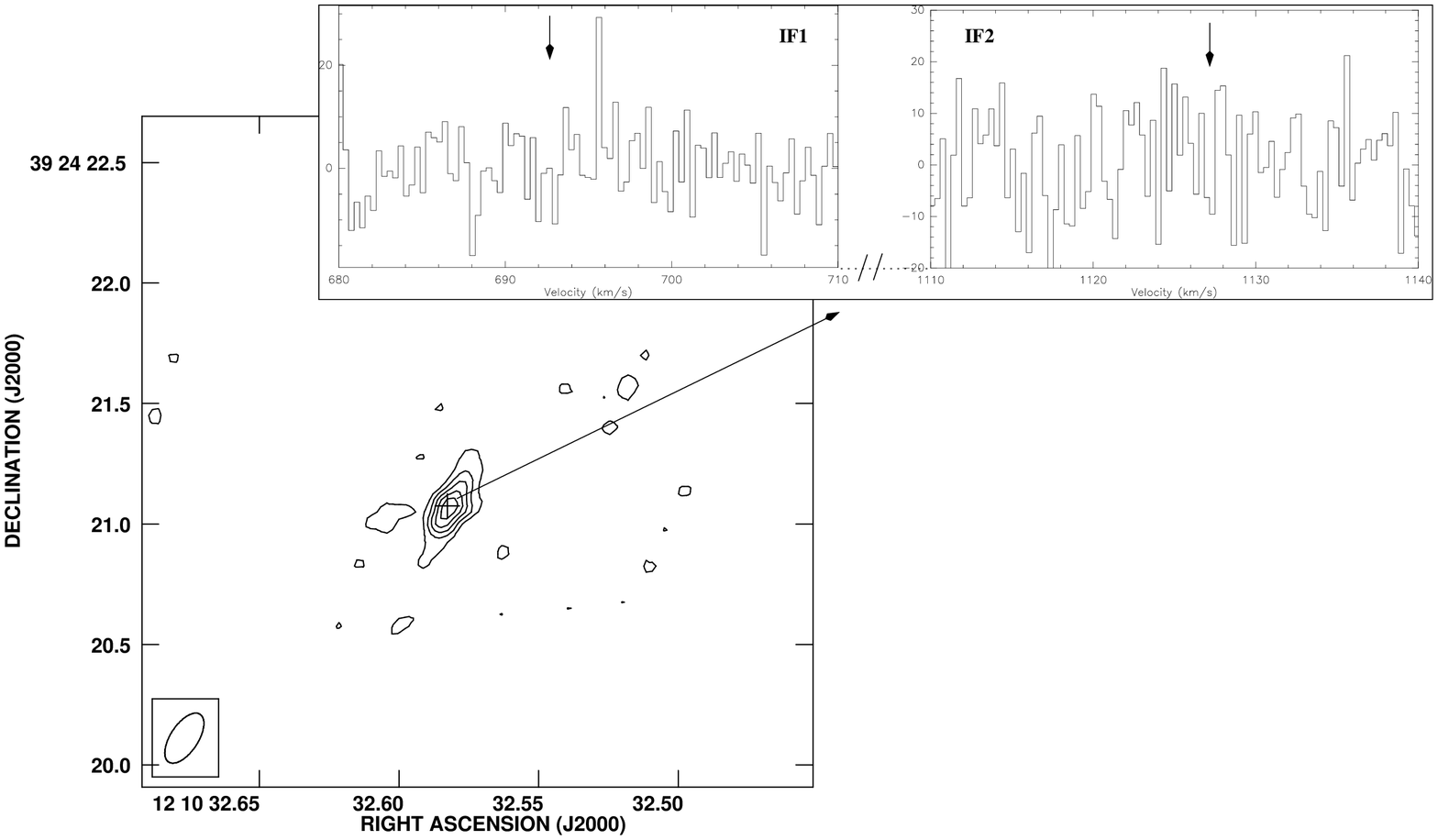}
\caption{
\textit{Lower-left panel}: naturally-weighted VLA DnA-array 22\,GHz radio continuum contour map of NGC\,4151 produced by using the line-free channels in the maser data cube. Contours: -1,1,2,3,4,5 $\times$ 1.7 mJy (1$\sigma$ rms = 0.6\,mJy\,beam$^{-1}$). \textit{Upper panels}: 22\,GHz spectra from the IF1 (left) and IF2 (right) VLA DnA-array data cube. The two IFs sample the velocity ranges where narrow water maser lines (their velocities are marked by arrows) were detected with the GBT by \citet{braatz04}. The systemic velocity of the galaxy is 995 \kms.} 
\label{fig:ngc4151_tot}
\end{figure*}

Our VLA DnA spectra (Fig.~\ref{fig:ngc4151_tot}, upper-right panel) indicate, in the first IF, the tentative presence of a narrow single-channel line at 695.6 \kms\ with a peak flux density of $\sim$ 30 mJy for a channel spacing of 0.33 \kms, while no emission is detected in the second IF. The position of the putative emission feature is $\alpha_{\rm J2000}$=12$^{\rm h}$10$^{\rm m}$32\fs58 and $\delta_{\rm J2000}$=39\degr24\arcmin21\farcs1. Assuming an upper limit for the uncertainty in the peak velocity of about one channel, 0.33 \kms, this feature, if real, is still offset by $\sim$ 3 \kms\ from the one detected by \citet{braatz04}. However, it may well belong to a group of variable features emitted in a small range of velocities. The indication of strong intrinsic variability in the NGC 4151 maser was indeed reported by \citet{braatz04}. 

The radio continuum contour map produced by using the line-free channels of the data cube is shown in the lower-left panel of Fig.~\ref{fig:ngc4151_tot}. A slightly-resolved source is detected at 22 GHz with position $\alpha_{\rm J2000}$=12$^{\rm h}$10$^{\rm m}$32\fs58$\pm$0\fs01 and $\delta_{\rm J2000}$=39\degr24\arcmin21\farcs1$\pm$0\farcs1 and a peak flux density of $\sim$ 15 mJy that is coincident in position, within the limits of accuracy ($\sim$ 0\farcs07; see Sect.~\ref{sect:posit} and Table~1), both with the maser spot and with the putative AGN of NGC~4151 (e.g. \citealt{carral90}; \citealt{mundell03}; \citealt{ulvestad05}).

% ___________________________________________________________
%

\section{Discussion}\label{disc}

\subsection{Individual galaxies}\label{indi}

\subsubsection{NGC\,3556}
\begin{figure*}
\includegraphics[width= 17 cm]{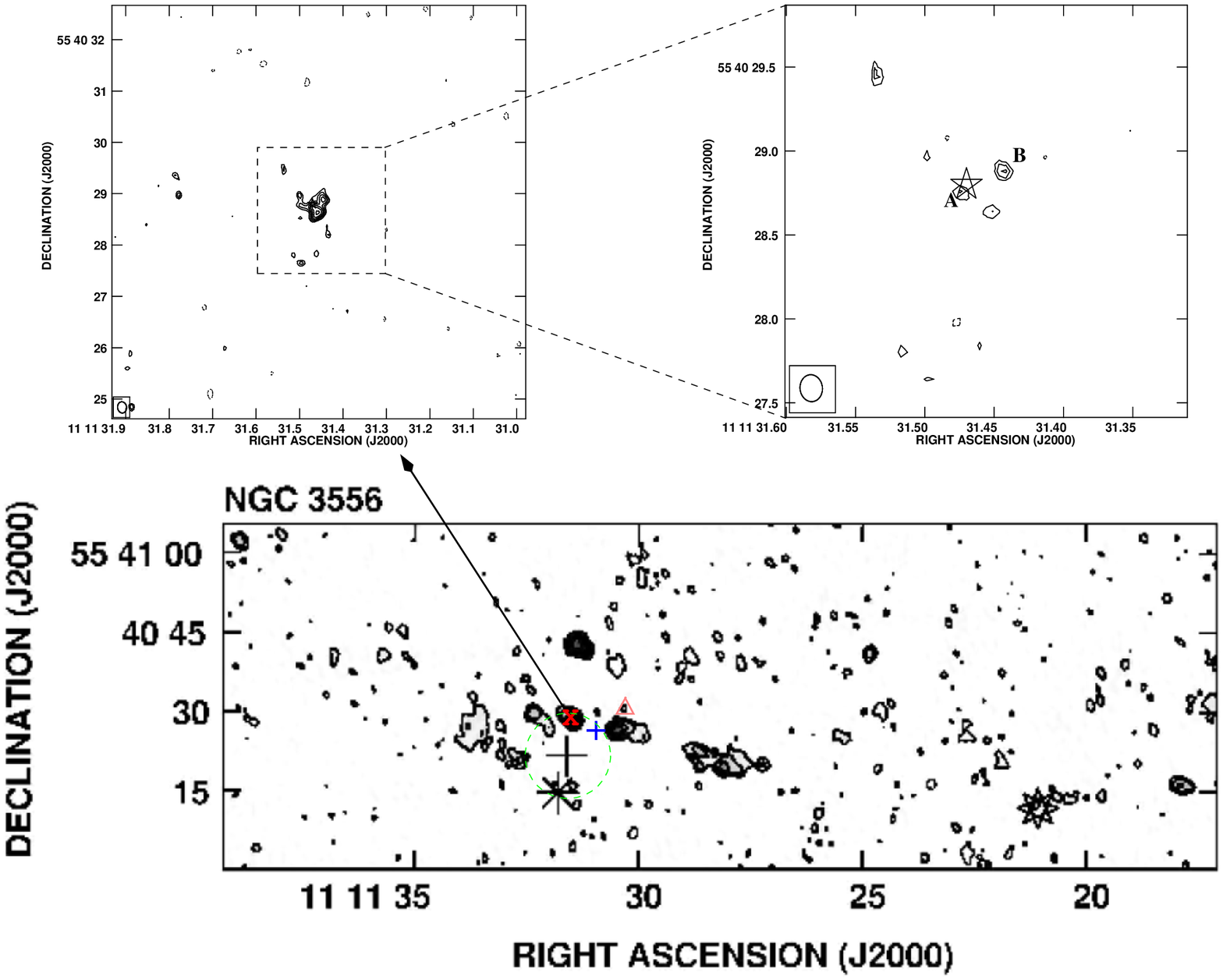}
\caption{\textit{Lower large panel}: A VLA A-array map of the continuum emission of NGC\,3556 at 20\,cm (ISE).
The asterisk identifies the optical center. The large black cross indicates the {\HI} kinematic center given by \cite{n3556king97} with
the relative error (dashed-line circle). The smaller blue cross near the upper-right corner of the green dashed circle denotes the 2${\rm \mu m}$ peak. The pink triangle marks the position of \textit{Chandra} 
source 35. The position of the water maser is indicated by the red X-mark. The multipointed star in the lower right edge indicates a position at which some supernovae have been detected (see ISE). 
\textit{Upper left panel}: naturally-weighted contour image of the only compact nuclear radio continuum source, among those of ISE, detected in NGC\,3556 at 1.4 GHz by MERLIN. Contours: -2.5,2.5,3,3.5,4,4.5,5 $\times$ 50 $\mu$Jy/beam (1$\sigma$ rms = 0.05\,mJy\,beam$^{-1}$).
\textit{Upper right panel}: uniformly-weighted 1.4 GHz MERLIN contour image of the compact source. The star marks the water maser position. Contours: -3,3,3.5,4 $\times$ 40 $\mu$Jy/beam (1$\sigma$ rms = 0.04\,mJy\,beam$^{-1}$).} 
\label{fig:n3556_global}
\end{figure*}

Since no continuum emission was detected in our 22-GHz map of NGC\,3556, we use maps published in the literature. The main maser spot is possibly associated with the strongest compact source (hereafter referred to as CS-3556) shown in the 1.4-GHz VLA map of ISE (Fig.~\ref{fig:n3556_global}, lower panel) although with a relatively large uncertainty (the relative position error is of 0\farcs1). Thus, the origin of the maser emission is closely related to the nature (AGN or star formation region) of CS-3556 and to the possible presence of an AGN.

While spectral index information on CS-3556 was not available, ISE concluded that the radio emission has a non-thermal
origin and is due to a collection ($\sim$\,10) of supernova remnants (SNRs) that cannot be resolved by the synthesized beam (although not mentioned by ISE, another possible explanation is that CS-3556 may actually be an individual bright SNR or radio supernova, like those found in M\,82).
%Their conclusion is mostly based on two elements: (1) the position of CS-3556 is offset ($\sim$\,7\arcsec\ $\equiv$\,400\,pc at the distance of NGC\,3556) from the putative nucleus (i.\,e. the {\HI} kinematic center). This argues against an AGN nature. (2) The 1.4 GHz radio power of CS-3556 is found to be intermediate between that of Cassiopeia A\footnote{The most luminous SNR in the Milky Way.} and that of the brightest SNR in M\,82 (ISE, their Table~3). 
 
%While, in our opinion, the large positional error of the {\HI} kinematic center (7\arcsec, \citealt{n3556king97}), that is of the same order as the nominal offset from CS-3556, does not allow to confidently rule out an association between this source and the nucleus, a relevant open question is: does NGC\,3556 presently host an AGN at all?
\cite{n3556wang03} found an AGN candidate among the discrete X-ray sources detected with \textit{Chandra} (hereafter labeled X-35).
 However, its position offset from both the {\HI} kinematic center and the 2${\rm \mu m}$ peak (14\arcsec\ $\equiv$\,815\,pc and 8\arcsec\ $\equiv$\,465\,pc, respectively; see Fig.~\ref{fig:n3556_global}) and X-ray luminosity ($\sim$\,10$^{39}$\,erg\,s$^{-1}$ in the 0.5--10\,keV band), just at the lower limit of the Eddington luminosity, indicates for X-35 a nature more consistent with a luminous X-ray binary than with a putative AGN. A clear signature of the presence of an AGN is then missing. Furthermore, the source CS-3556 is confidently offset from X-35 and the {\HI} peak (by 11\arcsec\ and 8\arcsec, respectively), and, in particular, from the 2${\rm \mu m}$ peak (the position which is known with an accuracy of $\sim$ 1\arcsec) by 5\arcsec. Hence, most likely, CS-3556 is indeed a group of SNRs as proposed by ISE, or an isolated particularly bright SNR like those found in the nuclear regions of other starburst galaxies, rather than the galactic nucleus. The fact that CS-3556 belongs to a chain of compact radio sources (Fig.~\ref{fig:n3556_global}) apparently tracing a partial ring or arm-like structure similar to those found in other galaxies and related to quasi-contemporary star formation triggered by a propagating density wave reinforces our conclusion. 

Very recently, we have obtained a deep 1.4-GHz MERLIN observation where we detected one continuum source in the nuclear region of NGC\,3556 confidently associated with CS-3556. The source is relatively weak (with a flux density of 0.25 mJy) and extended in the naturally-weighted image (Fig.~\ref{fig:n3556_global}, upper-left panel). When the source is imaged using uniform weighting, and hence, at a slightly higher resolution, the emission splits into two spots, labeled CS-3556/A and CS-3556/B (Fig.~\ref{fig:n3556_global}, upper-right panel), most likely hinting at being composed by two galactic SNR-like sources, as suggested by ISE, or compact \HII\ (C\HII) regions. Future spectral index studies of these two sources may discriminate between the thermal or non-thermal nature of the emission. In any case, this confirms with higher position accuracy the association between the maser and CS-3556 and, in particular, with CS-3556/A, the south-western of the two continuum sources. Then, if our scenario is correct, the confident association of the {\water} maser with CS-3556/A indicates that the kilomaser in NGC\,3556 is related to star formation like the majority of the weak extragalactic masers. 

%Our conclusion is reinforced further by a closer look at the Effelsberg spectra shown in HPT (their Fig.~3). The intensity of the maser in NGC\,3556 presents a variation in time, increasing from $\sim$\,40 to $\sim$\,60\,mJy between March 12 and March 15, 2002, to subsequently decrease to about 50\,mJy in the next 20 days. However, this maser variability hardly justifies 

The 50\% flux density decrease observed between the Effelsberg spectrum on September 27, 2002 (30\,mJy; HPT, their Fig.~3) and the VLA CnB profile (Fig.~1), obtained only a day after, hardly explainable in terms of variability, suggests that the maser feature observed with Effelsberg is a sum of a number of `weak' maser spots, that are resolved by the higher resolution of the VLA. Indeed, the channel maps (Fig.~\ref{fig:kntr_2}) indicate that the emission arises from more than one spot, even though only the strongest component is luminous enough to be confidently detected and to be present in more than one channel.
The possibility that we are observing a collection of weak masers, the brightest of which has a luminosity of $\sim$0.3\,{\solum}, favours once more the association with star formation activity.

%Why is blueshifted?
Interestingly, the velocity of the maser line ($V_{\rm LSR}$=738\,{\kms}) is redshifed by $\sim$\,40\,{\kms} w.r.t. the systemic velocity of the galaxy ($V_{\rm sys}$=699\,{\kms}, NED; $V_{\rm LSR} - V_{\rm HEL}$=\,+\,5.90\,{\kms} ). However, the maser line is located where the gas belonging to the {\HI} disk has velocities of about 680 \kms, which is blueshifted w.r.t. the systemic ones \citep{n3556king97}. Therefore, the discrepancy between the observed and expected velocity is of the order of 60\,{\kms}. 
%This, however, is not incompatible with gas motions in clouds in a complex environment like the one expected to host the masing gas (see also \citealt{surcis09} for a similar case in NGC\,3256). 
When excluding an association with the galactic nucleus, as discussed before, this discrepancy can be explained invoking gas motions in clouds (see, e.g. \citealt{surcis09} for NGC\,3256) or assuming that the masing gas belongs to a counter-rotating structure, an expanding supernova shell, or a starburst wind or outflow (see also \citealt{brunthal09} for a detailed analysis with respect to NGC\,253). With the data presently available for NGC~3556, it is impossible to discriminate between these possibilities.

\subsubsection{Arp\,299} 
The simultaneous detection of luminous water maser emission in different regions of Arp\,299 represents a unique case among the LIRG/ULIRG class of objects. Therefore, it provides an excellent opportunity to study origin and excitation mechanisms that form water masers in interacting galaxies and an important tool to understand the entire Arp\,299 merging system. 

As previously reported (Sect.~\ref{sect:a299}), one of the three main maser spots detected in Arp\,299 is coincident with the nuclear region of IC\,694. The higher spatial resolution map obtained from the VLA A-array indicates that the position of the maser is offset of $\sim$ 0\farcs2 ($\approx$ 40 pc at the distance of IC\,694) to the north-east w.r.t. that of the nuclear radio continuum peak at 8 GHz \citep{neff04}. However, due to the relatively low signal-to-noise ratio of the VLA-A array observations, this result requires further confirmation. Another masing spot is located within a distance of 0\farcs1 from the putative nucleus of NGC\,3690 (Fig.~\ref{fig:arp299_tot}). Then, a significant fraction of the emission is located at the overlapping region (the C+C'-complex), the majority of which associated with the subcomponent labeled C', although weaker emission is present also at the component C.

\begin{figure}
\includegraphics[width= 8.5 cm]{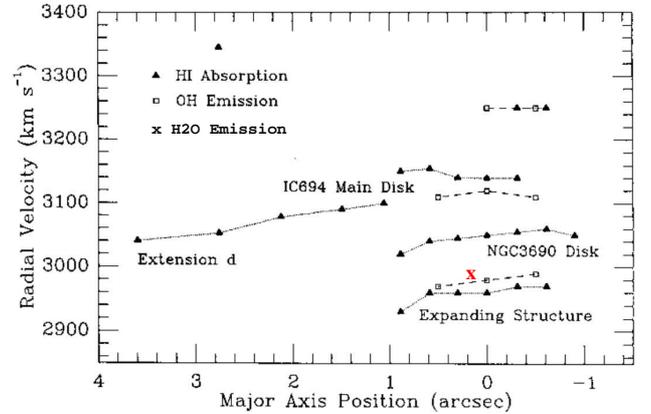}
\caption{This is a modified version of Fig.\,11 of BH90 showing the \HI\ and OH velocity components along the major axis (at position angle of $\sim$ 45\degr) near the nucleus of IC\,694. The red 'X' marks the position and velocity the water maser feature detected by us in IC\,694. The proximity to the expanding structure introduced  by BH90 and present in both \HI\ and OH data, makes an association between the water maser and this dynamical component likely.} 
\label{fig:ic694_vf}
\end{figure}

The velocities of the water masers are unusual when being compared with the \HI\ (BH90) and CO \citep{casoli99} velocity fields. In particular, the \water\ maser line associated with IC\,694 should have a velocity around 3140 \kms, while it is detected at $\sim$ 2990 \kms. The peak velocity ($\sim$ 3112 \kms) of the maser feature associated with the nucleus of NGC\,3690 is instead more consistent with, although slightly red-shifted to, that expected from that of other gas tracers at the same location ($\sim$ 3050 \kms). Furthermore, more than one velocity component is present in the C+C'-complex. 

The aforementioned discrepancies are, however, not surprising at all. The gas velocity field of Arp\,299, both in its atomic and molecular component, is very complex and cannot be described by a composition of two independent rotating disks representing the two main galaxies (e.g. BH90; \citealt{aalto97}). A simple picture to describe the behaviour of the \HI\ and OH gas in Arp\,299 has been proposed by BH90 where all the higher-velocity gas belongs to IC\,694 and that at lower velocity is related to NGC\,3690. In this scenario, gas from both galaxies reaches the other galaxy producing the apparent anomalies in the velocity field. Furthermore, gas from NGC\,3690 and IC\,694 have components also in C and C', respectively. While the maser found close to the nucleus of NGC\,3690 can be safely associated with this galaxy, given its precise overlapping in position with the nuclear source and small difference in velocity, the maser apparently associated with IC\,694, could instead be masing gas associated with the disk of NGC\,3690 possibly amplifying emission from the nuclear (star-forming or AGN) region in IC\,694. However, before drawing such a conclusion we should note that according to BH90 only the large scale velocity field is explained by the aforementioned simple model, while from a closer look in front of the nuclear region of IC\,694 several velocity components are present. In Fig.~\ref{fig:ic694_vf}, we show a modified version of their figure (Fig. 11 in BH90) where such a description is summarized and where we have included the position and velocity of our maser to determine with which of the velocity components it can be associated. Obviously, the maser is most likely associated with the expanding structure present in IC\,694. This structure, represented in both the HI absorption and OH maser emission data, is possibly a slab of material related to outflow phenomena due to superwinds produced in the AGN or regions with particularly-enhanced star formation activity. Strong \HI\ and OH outflows have been detected in several edge-on systems that share similarities to IC\,694. Hence, we think that the 2990 \kms\ feature is masing gas associated with IC\,694.

\begin{figure}
\includegraphics[width= 8.5 cm]{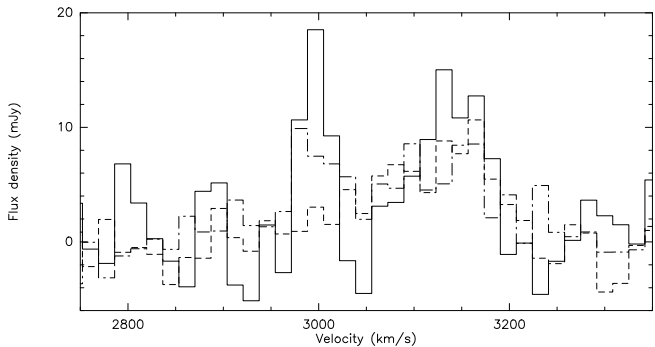}
\caption{Water maser spectra taken with the 100-m Effelsberg telescope on Nov.\ 2005, pointing towards putative nuclei of IC\,694 (dash-dotted line) and NGC\,3690 (dashed line), and on April 5, 2007 (4 days before the VLA D-array measurements) on a single pointing at intermediate distance between the nuclei (solid line). In this latter case, the Effelsberg 40$\arcsec$ beam covers the entire Arp\,299 system.} 
\label{fig:arp299_vn}
\end{figure}

As already mentioned, our results can help to determine the nature of the various objects constituting the Arp\,299 system. 

{\bf {IC\,694}} The debate on the nature of A, the putative nucleus of IC\,694, is still ongoing. Based on the compact nature of the radio emission some authors suggested the presence of an obscured AGN in A (e.g. \citealt{sargent91}; \citealt{lonsdale92}). From a comparison between the X-ray luminosity of A and B1, \citet{ballo04} deduced that both IC\,694 and NGC\,3690 host AGN activity. According to them, such X-ray intensity could not be accounted for by star formation alone. However, \citet{zezas03} argued against this interpretation. Many other authors indicate that the nuclear region of IC\,694 is exhibiting particularly strong and compact nuclear star formation activity with no evidence for an AGN (e.g. \citealt{alonso09} and references therein). 

One of the questions that arise when looking at the three water maser spectra of IC\,694 (that of Effelsberg, VLA-A, and VLA-D; Fig.~\ref{fig:arp299_tot}, upper-left inserted panel) is the difference in their peak fluxes. Indeed variability can be invoked to justify these flux changes. This seems to be supported by the comparison of the single-dish spectra taken with the 100-m Effelsberg telescope in November 2005 and in April 2007 (only 4 days before the VLA D-array measurements) shown in Fig.~\ref{fig:arp299_vn}, where the maser line associated with IC\,694 seems to experience an increase in flux density (the intensity of the maser in NGC\,3690 does not change much). Given, however, the different characteristics of the measurements (with the older observation being sensitive and pointing at the individual nuclei, while the more recent spectrum is noisier and targeting a middle point in the system), the variability scenario should be taken with caution. 

An alternative option may also be viable. The lower resolution spectra (the November-2005 Effelsberg and VLA-D ones) have similar line flux densities and profiles. The spectrum taken with the VLA A-array is instead quite different. Although (as reported in Sect.~2) these observations were affected by bad weather that increased the noise level, the 60 \% reduction in flux density w.r.t. to the other spectra has to be considered as real. If the maser in IC\,694 is associated with an AGN, as possibly suggested by its large interferometric brightness, ranging from 8 to 28 \solum\ depending on the VLA array used, we would not expect to resolve its emission, and hence, to significantly lose flux, when passing from the D to the A array. If instead, the observed line is actually a collection of numerous maser spots associated with the intense star-formation activity known to be present in IC\,694 (e.g., \citealt{alonso09} and \citealt{pereztorres09}), at the resolution of the VLA A-array, we may start to resolve this group of spots. We could then be 'left' with only the stronger ones, while the weaker may be below the detection threshold. Similar cases have been reported for a number of extragalactic kilomasers associated with star formation (\citealt{tarchi02}, for NGC\,2146; \citealt{surcis09}, for NGC\,3256; present work, for NGC\,3556). If this scenario is correct, the maser in IC\,694 would be associated with star formation rather than an AGN. An additional hint for the possible 'star-forming' nature of the maser is the 0\farcs2 offset derived between the radio continuum peak in source A and the maser location (see above). This, of course, does not provide a definite answer to the question on the presence of AGN activity, especially since OH (mega)maser emission was found in source A (and only there in the entire system). Very recently, \citet{ulvestad09}, has reported on a study of the radio luminosities of the 25 continuum compact sources detected in the nuclear region of IC\,694 with sensitive VLBI observations, concluding that the possibility that one of these sources is associated with an AGN cannot be ruled out.

For IC\,694, we also compared the positions of the emission produced by the two maser species. The VLA study by BH90 indicates that the OH emission is coincident with the central peak of radio emission in IC\,694 but the emission is extended. Furthermore, OH pumping requires that the molecular gas is located close to the nuclear FIR region. The spatial coincidence between the OH maser and the radio continuum peak has been confirmed at higher resolution also with MERLIN (Polatidis \& Aalto 2000, 2001) and, tentatively, with the EVN \citep{kloeckner02}. The location of the water maser emission is also very close to the nuclear region although a small offset w.r.t. the radio continuum  seems to be present (see previous paragraphs). Since the OH maser absolute position is not reported in \citet{polatidis00}, \citet{polatidis01}, and \citet{kloeckner02} and the alignment accuracy between our VLA A-array map and that of BH90 is limited to $\sim$ 0\farcs2, the possible coincidence between the two masing regions cannot be ruled out. In any case, our result provides first evidence for strong maser emission from both molecules.

{\bf {NGC\,3690}} Several attempts have been made to unveil the presence of AGN activity within this galaxy. \citet{dellaceca02}, \citet{zezas03}, and \citet{ballo04} using X-ray observations, \citet{garcia06} through optical measurements, and \citet{gallais04} and \citet{alonso09} with studies at infrared wavelengths, all strongly support the hypothesis that B1 is the true nucleus of NGC\,3690 hosting an AGN. The detection of our water maser confirms this conclusion. The maser spot's location is coincident, within an accuracy of 0\farcs1 ($\approx$ 20 pc at a 42 Mpc distance), with the putative nucleus of the galaxy (B1). Furthermore, water masers found in association with AGN have typically luminosities $>$ 10 \solum\ and are forming a collection of spots closely grouped around the nucleus, thus only being seen as a single feature at the angular scale of the VLA. The maser in NGC\,3690 follows these characteristics, has an isotropic luminosity of $\sim$ 20 \solum, is spatially unresolved with the VLA A-array (100\,mas resolution) and does not show order of magnitude variations in its overall integrated flux density. The presence of an AGN in NGC\,3690 is also considered a possibility by very recent radio continuum VLBI observations performed by \citet{ulvestad09}.  

{\bf {C+C'-complex}} Another truly relevant result of our VLA observations is the detection of emission in the C+C'-complex. Given its noticeable properties, e.g. the high molecular mass and the conspicuous NIR emission, it has been suggested that this complex were a satellite galaxy taking part in the merger event (e.g. \citealt{casoli89}; \citealt{telesco85}; \citealt{nakagawa89}). However, BH90 concluded that it is not necessary to invoke the existence of a third galaxy in the system to justify the region C's properties. Our finding, the presence of \water\ kilomaser emission associated with the C-complex both in the subcomponents C' and, at a weaker level, C, favors vigorous star formation and does not provide any strong hint for the presence of an active nuclear core. The youth and intensity of this burst as deduced from its exceptionally strong 22-GHz emission and the presence of a substantial population of (ultra-)compact/ultra-dense (UC/UD) \HII\ regions (e.g., \citealt{alonso09}) is remarkable. A similar case has been recently reported by \citet{DBJ08} for the Antennae system (NGC\,4038/4039), where water kilomaser emission of $\sim$ 8 \solum\ has been detected in the ``interaction region'' (IAR) offset from the nuclei of the two galaxies taking part in the merger. The emission arises from one spatially-unresolved spot separated in velocity into two distinct features. The water kilomaser emission is similarly luminous in the Antennae and the C-complex of Arp\,299. This further corroborates similar physical conditions, as already suggested earlier by common properties like enhanced star formation, strong (N)IR emission, and the presence of several UD\,\HII\ regions \citep{DBJ08}.

The Arp\,299 and Antennae merging systems themselves are, however, quite different. Assuming a distance of 42 Mpc for Arp\,299 (see Sect.~\ref{sect:a299}) and of 21 Mpc for the Antennae (\citealt{DBJ08}), the IR luminosity of the former (L$_{\rm IR}$ $\approx$ 5$\times10^{11}$ \solum; \citealt{alonso00}) is 5 times that of the latter (L$_{\rm IR}$ $\approx$ $10^{11}$ \solum; \citealt{sanders03}). Furthermore, the linear distance between the two galactic nuclei in the systems (IC\,694/NGC\,3690 and NGC\,4038/4039) derived from interferometric radio maps are 5 \citep{neff04} and 7 kpc \citep{neff00}, respectively. In general, the merger in Arp\,299 is more advanced than in the Antennae (e.g. \citealt{sargent91}; \citealt{schulz07}). Since galaxy mergers are a very efficient means of fueling AGN (e.g., \citealt{combes01}), the different stage of the mergers may account for the presence of AGN activity in Arp\,299 that is not detected in the Antennae. At the same time, since advanced mergers may result in burying AGN with large concentrations of obscuring material, the large nuclear obscuration would also favor the occurence of very luminous maser emission. This would further explain our water maser detection in the nuclei of the two galaxies of Arp\,299. No maser emission has been so far reported in either one of the two nuclei of the Antennae. Indeed, the nucleus of NGC\,4038 has been searched for water maser emission down to a luminosity detection threshold of $\sim$ 1 \solum\ (HPT), while no mention in the literature is made for a similar search in the nucleus of NGC\,4039.

\subsubsection{NGC\,4151}
The maser line tentatively detected in IF1 comes from an unresolved spot coincident in position with the continuum peak detected in the line-free channel map. From its location this peak can be identified as the sub-component E of the radio knot C4 (labels  are from \citealt{mundell03} and references therein). While the debate on the exact location of the AGN in NGC\,4151 is still ongoing, the most recent publications indicate that D, and in particular sub-component D3, is the most promising candidate based on the location and distribution of the \HI\ absorption, the structure of the radio continuum (\citealt{mundell03}), and the high brightness temperature of the source (\citealt{ulvestad05}). According to \cite{mundell03}, the source E is instead the brightest knot of the radio counterjet that is located behind the \HI\ absorbing layer. If this is the case, the kilomaser in NGC\,4151 may be produced by an interaction between the radio jet and the interstellar medium (ISM). Indeed, indications of such an interaction between the radio jet and the clumpy ISM of NGC\,4151 are reported by \cite{mundell03}. NGC\,4151 may then represent the second case, after that of M\,51 (\citealt{m51}), of a nuclear kilomaser. Similarities do exist between the two cases like the presence of redshifted and blushifted emission in the single-dish spectra, although the maser features in M\,51 are broader than those in NGC\,4151, and hence, more consistent with the linewidths found for the megamaser sources associated with radio jets. The strength, narrowness, and strong variability of the maser features favor instead an association with star formation activity that, with the present data, cannot be excluded.

\subsection{Kilomasers revisited}\label{kilo}

In Table~\ref{kilomasers}, we have summarized relevant results related to extragalactic {\water} kilomasers, in the context of their association with AGN, SNRs, or {\HII} regions. We have a total of 32 kilomaser sources in 24 galaxies. Half (16) of these 32 kilomasers (labeled with ``SF'') are confidently associated with star formation activity being either clearly off-nuclear or, like NGC\,253\,\water--1 and NGC\,3556, coincident in position with sources in the nuclear region produced by star formation phenomena. For the 8 sources labeled with ``SF(?)'', the association with star formation is instead (only) the most likely option since it is based on qualitative considerations on the nuclear position of the maser emission, the absence of clear evidence for AGN activity in the host galaxy, and/or the shape of the maser line profile (see e.g., \cite{castangia08} for NGC~520, \cite{surcis09} for NGC~3620).

\begin{figure}
\includegraphics[width= 8.5 cm]{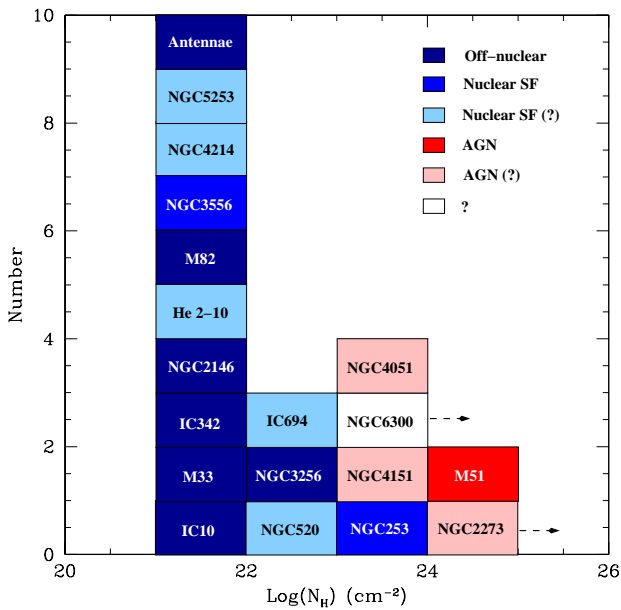}
\caption{Histogram showing the number of \water\ kilomaser galaxies as a function of their nuclear column density derived from X-ray spectroscopy. Due to the higher angular resolution, Chandra data are taken whenever possible (with the exception of NGC\,4051 for which the XMM-derived column density value was preferred to that of Chandra because of the large uncertainties reported). Otherwise, the most recent measurements are used (see Table~\ref{colden}). The arrows for NGC\,2273 and NGC\,6300 indicate lower limits to the column densities. Each color indicates the association of the water maser emission with either star formation or AGN activity, and the degree of confidence into such an association.} 
\label{fig:kilo}
\end{figure}

For other 4 kilomasers (in NGC\,2273, NGC\,4051, NGC\,4151, and NGC\,4293, labeled with ``AGN?'') the origin of the maser emission has been instead plausibly related to AGN activity (although, also in this case, a validation is necessary), mainly because the nucleus of the galaxy hosting the maser emission is an AGN and/or the maser line profile resembles more that of the AGN-associated ones (see, e.g., \cite{n4051} for NGC~4051; the present work for NGC~4151). Among the remaining sources, for 2 of them (in NGC\,1106 and NGC\,4527) both associations, AGN or star-formation, are equally plausible (\citealt{BG08}), while for the kilomaser in NGC~6300 the nature of the maser emission, so far, has never been discussed. Noticeably, the only kilomaser ever reported that is undoubtedly associated with an AGN (labeled with ``AGN'') and, in particular, with the nuclear radio jet, is that in M~51 (\citealt{m51}).

Hence, the question arises: is there a family of nuclear, AGN-associated, kilomasers? Are there intrinsic differences in the host galaxy that, in absence of studies at high ($<$ 1\arcsec) spatial resolution, can allow us to discriminate between kilomasers associated with star formation and those associated with AGN activity? A possible suggestion was briefly introduced by \cite{zhang06} and, more recently, by \cite{greenhill08} in the larger framework of the association between maser phenomenon and nuclear X-ray column density. Their general conclusion was that the distributions of kilomasers and megamasers are different, with the latter being associated with galaxies with higher nuclear column densities (half of the megamasers in their sample are Compton-thick, viz. $N_{\rm H} > 10^{24}\,{\rm cm}^{-2}$). In particular, \cite{zhang06} found that, for kilomasers, two groups were apparently present, one with nuclear column densities of $10^{21-22}\,{\rm {cm}}{-2}$ and another one with nuclear column densities of $> 10^{23.3}\,{\rm {cm}}{-2}$. Furthermore, an indication was put forward that the few nuclear kilomasers have an average column density indistinguishable from those of the entire megamaser sample. In Fig.~\ref{fig:kilo}, we show an updated plot of that of \cite{zhang06} (their figure 6a), with the number of \water\ kilomasers as a function of nuclear column density derived from X-ray spectroscopy for all 19 kilomaser galaxies (\citealt{zhang06}: 12 galaxies) for which such information was available. From Fig.~\ref{fig:kilo}, it is clear that the lower average nuclear column density of kilomasers w.r.t. that of megamasers is confirmed, while there is no evidence for the 'bimodal' distribution of nuclear column densities within kilomasers. The value for the nuclear column densities is a continuous function that spans a range of values between $10^{21.08}\,{\rm {cm}}^{-2}$ (M\,82) and $10^{24.75}\,{\rm {cm}}^{-2}$ (M\,51; although the values for NGC\,6300 and NGC\,2273 are only lower limits). Indeed, as qualitatively suggested by \cite{zhang06}, the kilomasers associated, or most likely associated, with AGN activity are found in galaxies with larger nuclear column densities ($10^{23}\,{\rm {cm}}{-2}< N_{\rm H} < 10^{25}\,{\rm {cm}}{-2}$), differently from those with star forming origin that are found in galaxies with nuclear column densities $< 10^{23}\,{\rm {cm}}{-2}$. The only exception is represented by the maser in NGC\,253 that, despite being quite confidently associated with star formation (\citealt{hofner06}; \citealt{brunthal09}), is found in a galaxy with relatively high column density ($2\times10^{23}\,{\rm {cm}}{-2}$). According to this view, kilomasers like those in NGC\,6300 or NGC~2273, for which the nature of the emission is still uncertain, could be speculatively associated with AGN activity because of the large nuclear column density. Interestingly, very recent results indicate that the latter maser source is indeed associated with an accretion disk around the central engine of the galaxy \citep{greene10}. Unfortunately, for five kilomaser galaxies (NGC\,1106, NGC\,3359, NGC\,3620, NGC\,4293, and NGC\,4527), no information on column densities could be obtained. Future observations of these targets by $XMM-Newton$ and $Chandra$ would be desirable.

If our picture is correct, then the presence of a family of nuclear kilomasers associated with AGN activity is possible. One of the distinguishing characteristics may indeed be the higher obscuration present around nuclear engines that we infer from the nuclear X-ray column densities. Since our discussion only involves nuclear X-ray column densities, we cannot (and do not want to) investigate the relation between the kilomaser phenomenon and the X-ray absorption in off-nuclear kilomaser regions. However, these detections originate from single-dish searches for water maser emission that, for more distant objects, cover the entire body of the target galaxy or, for the nearer ones, most of the galactic disk extent, including the nuclear regions. Hence, the lack of nuclear maser emission in these galaxies having relatively small nuclear X-ray column densities reinforces our supposition.

According to our results, the typical distinction between two classes of extragalactic water maser sources, kilo- and mega-masers, based on a sharp isotropic luminosity edge, traditionally set to 10 \solum\, should, in our opinion, be revised. While it is quite safe to state that the upper end of the water maser luminosity function is populated exclusively by maser sources associated with AGN activity, at lower luminosities a mixture of masers associated with either star formation or AGN activity seems to be present. In particular, since maser isotropic luminosities, especially for those AGN-related, may be strongly dependent on orientation effects, the existence of relatively low-luminous AGN-related masers may be justified. Hence, if necessary, a more proper way to distinguish between water masers should be related to the origin of the emission associated to either one of the aforementioned forms of activity.  
%Of course, the possibility exists that the relation between nuclear kilomaser and higher column densities can just be a selection effect due to the location of the maser emission itself. That is, if the maser is nuclear, whatever its nature is, then the obscuration is higher. When the maser is off-nuclear is, by definition, of star-forming origin and the column density in a off-nuclear region is expected to be lower.

\begin{table*}
%{\scriptsize{
\centering
\caption{Properties of extragalactic 22\,GHz {\water} kilomasers and their potential association with AGN, SNRs or {\HII} regions.}
\label{kilomasers}
%\vspace{0.2cm}
\begin{minipage}[t]{\textwidth}
\renewcommand{\footnoterule}{}
%\hspace{-0.5cm}
\begin{tabular}{lrcccl}
\hline\hline

Source & Distance\footnote{The isotropic luminosity is that of individual maser spots when the emission is spatially resolved. Otherwise, the full luminosity is taken from a single-dish measurement. For the sources not studied in the present work, distances and luminosities are taken from HPT.} & {\wlum}$^{a}$ & Association\footnote{The reported labels identify the maser association: ``SF'' and ``AGN'' indicate that interferometric measurements have confidently associated the maser emission with either star formation activity or an active galactic nucleus. A '?' sign is added to the favourite option, if there exist still some doubts. For unexplored sources, only a '?' sign is given.} & Notes  & References \\
       &  (Mpc)   & {\solum} &             &  & \\

\hline

IC\,10--NW              & 1.2   & 0.02 & SF             & Molecular cloud    &        (\citealt{ic10becker93})  \\ 
IC\,10--SE              &       & 0.15 & SF             & {\HII} region      &        (\citealt{ic10becker93})  \\ 
NGC\,253 {\water}--1   & 3.0   & 0.15 & SF             & SNR                &        (\citealt{hofner06};\citealt{brunthal09}) \\
NGC\,253 {\water}--2   &       & 0.02 & SF             & Star forming region&        (\citealt{n253})           \\
NGC\,520               & 30.0  & 1.00 & SF(?)          & SNR                &        (\citealt{castangia08}     \\
M\,33-IC\,133          & 0.7   & 0.32 & SF             & {\HII} region      &        (\citealt{m33})            \\
M\,33-19               &       & 0.03 & SF             & {\HII} region      &        (\citealt{m33/19})         \\
M\,33-50              &       & 0.02 & SF             & {\HII} region      &        (\citealt{m33/50},\citealt{brunthal06})         \\
NGC\,1106              & 57.8  & 8.0  & ?              &                    &        (\citealt{BG08})          \\
IC\,342                & 2.0   & 0.01 & SF             & Star forming region&        (\citealt{ic342})         \\
NGC\,2146--A            & 14.5  & 0.50 & SF             & UC{\HII} region    &        (\citealt{tarchi02})         \\
NGC\,2146--B            &       & 1.20 & SF             & UC{\HII} region    &        (\citealt{tarchi02})          \\
NGC\,2273              & 24.5  & 7.00 & AGN(?)         & Compton-thick      &        (\citealt{zhang06}) \\
He 2-10                & 10.5  & 0.68 & SF(?)          &                    &        (\citealt{DBJ08})              \\
M\,82-NE                 & 3.7   & 0.02 & SF             & {\HII} regions     &        (\citealt{hagiwara07})       \\
M\,82-SW                 & 3.7   & 0.10 & SF             & {\HII} regions     &        (\citealt{hagiwara07})       \\
NGC\,3256--N            & 37.4  & 1.00 & SF             &                    &        (\citealt{surcis09})          \\
NGC\,3256--S            &       & 2.40 & SF             &                    &        (\citealt{surcis09})         \\
NGC\,3359              & 13.5  &  0.7 & SF(?)          &                    &        (\citealt{BG08})              \\
NGC\,3556              & 12.0  & 1.00 & SF             & SNR                &        present work        \\
NGC\,3620              & 22.4  & 2.10 & SF(?)          &                    &        (\citealt{surcis09})          \\
Arp\,299--A             & 42.0  & 7.9  &  SF(?)         &                    &        present work        \\
Arp\,299--C'            &       & 3.4  &  SF            &                    &        present work                  \\
Antennae               & 20.0  &  8.2 & SF(?)          &                    &        (\citealt{DBJ08})             \\
NGC\,4051              & 10.0  & 2.00 & AGN(?)         &                    &        (\citealt{n4051})        \\
NGC\,4151              & 13.5  & 0.63 & AGN(?)         &                    &        (\citealt{braatz04}; present work) \\
NGC\,4214              & 2.9   &0.028 & SF(?)          &                    &        (\citealt{DBJ08})              \\
NGC\,4293              & 11.9  & 1.26--5.01 & AGN(?)   & jet(?)             &        (\citealt{kondratko06})       \\
NGC\,4527              & 23.1  & 4.0  & (?)         &                    &        (\citealt{BG08})              \\
M\,51                  & 10.0  & 0.63 & AGN            & Radio jet          &        (\citealt{m51})       \\
NGC\,5253              &  3.3  &0.021 & SF(?)          &                    &        (\citealt{DBJ08})            \\
NGC\,6300              & 15.0  & 3.16 & ?              &                    &        (\citealt{n6300})      \\
%NGC\,6946              & 0.6   & ?    & ?              &                    &        (\citealt{claussen1})     \\

\hline

\end{tabular}
\end{minipage}
%}}
\end{table*}

\begin{table*}
%{\scriptsize{
\begin{center}
\caption{Nuclear X-ray absorbing column densities of extragalactic H$_{2}$O kilomaser galaxies$^{*}$. This table is an updated version of that presented by \citet{zhang06}.}
\label{colden}
%\vspace{0.2cm}
\begin{minipage}[t]{\textwidth}
\renewcommand{\footnoterule}{}
%\hspace{-0.5cm}
\begin{tabular}{llclcl}
\hline\hline

 Source &  Type$^{\rm a}$ & Telescope$^{\rm b}$ &  Epoch$^{\rm c}$   & $N_{\rm H}$            & References\\
        &                 &                     &                    & (10$^{23}$\,cm$^{-2}$) &         \\
\hline\vspace{0.1cm}
IC\,10                       & X-1               & C  & 2003.03.12 & $ 0.06^{+0.002}_{-0.0008}$ & \citet{bauer04} \\\vspace{0.1cm}
NGC\,253                     & SBG               & B  & 1996.11.29 & $ 0.12_{-0.04}^{+0.03}$    & \citet{cappi99}  \\\vspace{0.1cm}
                             &                   & C  & 1999.12.16 & $ 2.0^{+1.3}_{-0.9}$       & \citet{weaver02} \\\vspace{0.1cm}
NGC\,520             & X-12              & C  & 2003.01.29 & 0.23                       & \citet{read05}   \\\vspace{0.1cm}
                             &                   & R  & 2003.01.29 & 0.029        & \citet{henriksen99}   \\\vspace{0.1cm}
M\,33                        &                   & A  & 1993.07    & $0.016^{+0.002}_{-0.002}$  & \citet{takano94}       \\\vspace{0.1cm}
                             & X-8               & X  & 2000--2002 & $0.019^{+0.005}_{-0.005}$  &  \citet{foschini04}  \\\vspace{0.1cm}
NGC\,1106            &  --                 & --   &   --         &       --             &     --   \\\vspace{0.1cm}
IC\,342                      &                   & A  & 2000.02.24 & $0.084^{+0.042}_{-0.042}$  &  \citet{kubota02} \\\vspace{0.1cm}
                             & X-21              & X  & 2001.02.11 & $0.087^{+0.013}_{-0.025}$  & \citet{kong03}\\\vspace{0.1cm}
NGC\,2146                    & SBG               & A  & 1997.03.26 & $0.021^{+0.022}_{-0.011}$  & \citet{dellaceca99} \\\vspace{0.1cm}
NGC\,2273                    & S2, C             & B  & 1997.02.12 & $>100$                & \citet{maiolino98}\\\vspace{0.1cm}
                             &                   & X  & 2003.09.05 & $\geq 18$             & \citet{guainazzi05} \\\vspace{0.1cm} 
He\,2-10            & BCD               & R  & 1992.05.14 & $0.0084-0.047$         & \citet{mendez99}\\\vspace{0.1cm}
                             &                   & C  & ...        & $0.0123^{+0.002}_{-0.002}$ & \citet{ott05}\\\vspace{0.1cm}
M\,82$^{\rm d)}$             & SBG               & B  & 1997.12.06 & $0.058^{+0.014}_{-0.015}$  & \citet{cappi99} \\\vspace{0.1cm}
                             &                   & X  & 2001.05.06 & $0.017^{+0.018}_{-0.016}$  & \citet{stevens03} \\\vspace{0.1cm}
                             &                   & C  & 2002       & 0.012                      & \citet{strickland04}\\\vspace{0.1cm} 
NGC\,3256            &                   & X  & 2001.12.15 & 0.015                      & \citet{jenkins04}       \\\vspace{0.1cm}
                             &  X-7 (N)          & C  & 2000.01.05 & 0.01             & \citet{lira02}       \\\vspace{0.1cm}
                             &  X-8 (S)          & C  & 2000.01.05 & 0.5              & \citet{lira02} \\\vspace{0.1cm}
NGC\,3359           & --                  & --   &  --          &    --                & --           \\\vspace{0.1cm}
NGC\,3556                    & X-35              & C  & 2001.09.08 & $0.03^{+0.017}_{-0.009}$   & \citet{wang03}             \\\vspace{0.1cm}  
NGC\,3620            & --                  & --   &    --        &     --               &       --     \\\vspace{0.1cm}
Arp\,299 (A; IC\,694)        & X-16              & C  &            & $0.119^{+0.102}_{-0.074}$  & \citet{zezas03} \\\vspace{0.1cm}
Arp\,299 (C'; overlap)       & X-8               & C  &            & ???               & \citet{zezas03}        \\\vspace{0.1cm}
Antennae             & X-25 (NGC\,4038)  & C  & 1999.12.01 & $0.107^{+0.036}_{-0.042}$  & \citet{zezas02}   \\\vspace{0.1cm}
                             & X-29 (NGC\,4039)  & C  & 1999.12.01 & $0.014$               & \citet{zezas02} \\\vspace{0.1cm}   
NGC\,4051                    & S1/1.5            & A  & 1993.04.25 & $0.045_{-0.035}^{+0.035}$  & \citet{george98} \\\vspace{0.1cm}
                             &                   & C  & 2000.04.25 & $\sim 0.01$         & \citet{collinge01} \\\vspace{0.1cm}
                             &                   & X  & 2001.05.16 & $\sim 3.6$          & \citet{pounds04}       \\\vspace{0.1cm}
NGC\,4151                    & S1.5              & A  & 1993.12.07 & $0.635_{-0.035}^{+0.041}$  & \citet{george98} \\\vspace{0.1cm}
                             &                   & A,B& 1999-2000  & $2.0^{+0.1}_{-0.1}$        &  \citet{schurch02}       \\\vspace{0.1cm}
                             &                   & C  & ...        & $1.5$                    & Wang et al. (2010)     \\\vspace{0.1cm}
NGC\,4214           & BCD               & C  & ...        & $0.101_{-0.011}^{+0.014}$  & \citet{ott05} \\\vspace{0.1cm}
NGC\,4293                    & LINER             & R  & ...        & see Table 3                & HEASARC       \\\vspace{0.1cm} 
NGC\,4527            & --                  & --   &  --          &    --                &    --         \\\vspace{0.1cm}
M\,51                        & S2, C             & A  & 1993.05.11 & $7.5^{+2.5}_{-2.5}$ & \citet{bassani99},\citet{terashima98} \\\vspace{0.1cm}
                             &                   & B  & 2000.01.18 & $56^{+40}_{-16}$      &  \citet{fukazawa01}\\\vspace{0.1cm}
NGC\,5253            & Dwarf SBG         & C  &  ...   & $0.101_{-0.003}^{+0.003}$  & \citet{ott05} \\\vspace{0.1cm}
NGC\,6300                    & S2, C?            & RX & 1997.02    & $>100$         & \citet{leighly99}\\\vspace{0.1cm}
                             &                   & B  & 1999.08    & $2.10^{+0.10}_{-0.10}$     & \citet{guainazzi02} \\\vspace{0.1cm}
                             &                   & X  & 2001.03.02 & $2.15^{+0.08}_{-0.09}$     & \citet{matsumoto04}\\
%{\it {NGC\,6946}}            & X-45              & C  & 2001.09.07 & $0.06^{+0.051}_{-0.066}$   & Hol03              \\
\hline

\end{tabular}
\end{minipage}
\end{center}
$^{*}$ {\it Chandra} and {\it XMM-Newton} results are given whenever possible. For sources not observed by these satellites
most recent (but usually older) results  are quoted
\ \ \ \ \

{\em a)} Type of nuclear activity. In the absence of a starburst and a prominent nuclear source, the X-ray source (e.g. X-1 or X-8) is 
given. BCD: Blue Compact Dwarf; SBG: StarBurst Galaxy; S2: Seyfert 2; LINER: Low-Ionization Nuclear Emission Line Region; ULIRG: UltraLuminous InfraRed 
Galaxy; FRII: Fanarov-Riley Type II galaxy; C: Compton-thick, i.e. $N_{H}$$\ga$10$^{24}{\rm {cm}}{-2}$; C?: {possibly varying} between 
Compton-thick and Compton-thin

{\em b)} A: {\it ASCA}; B: {\it BeppoSax}; C: {\it Chandra}; R: {\it ROSAT}; RX: {\it RXTE}; X: {\it XMM-Newton} 

{\em c)} Year, month and day

{\em d)} For NGC\,3034 (M\,82), lacking a well defined nucleus, the {\it Chandra} data refer to the diffuse
halo component.

\end{table*}

\begin{figure}
\includegraphics[width= 9.2 cm]{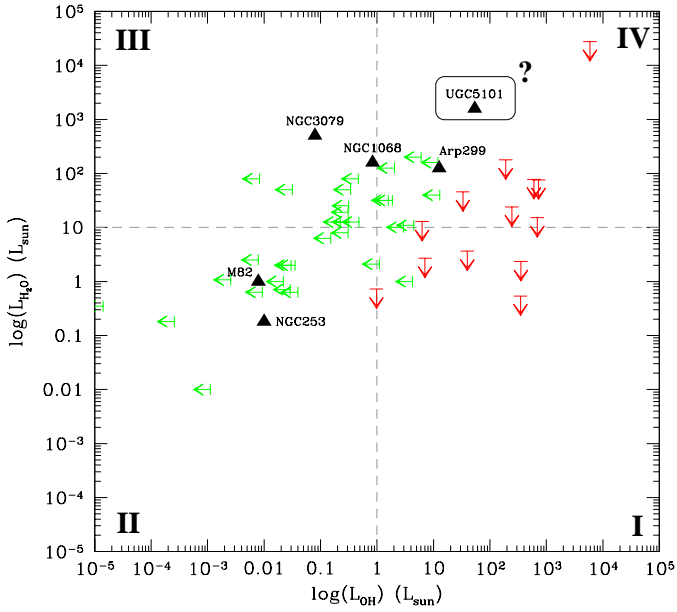}
\caption{OH vs. \water \, maser luminosity including all sources where masers from both species have been searched for. Green arrows represent the galaxies hosting only \water \, maser emission, red arrows are those sources hosting only OH maser emission, and filled triangles are galaxies where both \water \, and OH masers have been detected. Upper limits for the undetected maser species are set to the detection luminosity threshold ($3\sigma$) of the survey in that maser transition. The dashed lines indicate the separation between kilomaser and megamaser regimes as found in the literature (for \water \, masers, see Sect.\ 1; for OH masers, see \citealt{hen90}). The question mark associated with the position in the plot of UGC\,5101 is explained in Sect.~\ref{relation}.} 
\label{fig:rel}
\end{figure}

\subsection{On the \water/OH relation}\label{relation}

Table~\ref{waterandoh} (see the electronic edition of the Journal) lists all sources where maser emission from either \water\ or OH, or both, has been, so far, reported in the literature\footnote{Additional 50 unpublished galaxies, where 22\,GHz water emission has been recently detected, are listed in the `Megamaser Cosmology project' homepage. However, no luminosity information has been, so far, reported, and hence, these sources are not included in our list}. Out of the 173 sources of Table~\ref{waterandoh} (see the electronic edition of the Journal), 85 host water maser emission and 94 have OH maser emission detected. Six sources overlap between the two groups since they show maser emission from both molecular species. Unfortunately, not all sources that host maser emission from either one of the maser species have been also searched for emission from the other molecule.     
In Fig.~\ref{fig:rel}, we show a plot with the water and OH maser isotropic luminosities (measured values for detections or 3$\sigma$ upper limits in case of non detections), mainly derived from single-dish measurements, for those galaxies (51) searched for maser emission from both species. The plot can be summarized as follows:

\begin{itemize}
\item 32 galaxies show only water maser emission
\item 13 galaxies show only OH maser emission
\item no OH megamaser shows also \water\ kilomaser emission (quadrant I)
\item 6 sources show maser emission from both molecules. In particular, the two well-known starburst galaxies NGC\,253 and M\,82 show kilomaser emission from \water\ and OH (quadrant II), the two water megamaser galaxies NGC\,3079 and NGC\,1068 also host an OH kilomaser (quadrant III), and Arp\,299 and UGC\,5101 (doubts, however, on the result for this galaxy are described later in this section) show megamaser emission from both molecules (quadrant IV).
\end{itemize} 

Being aware that statistics may be affected by the different and/or insufficient sensitivities of maser searches and by possible variations in the maser features, in the following, we try to provide a possible interpretation of these results.
Despite our warning on using a strict isotropic luminosity threshold to infer the origin of water maser emission (see previous section), for consistency with past works (see e.g. \citealt{surcis09}, and references therein), in the following, we necessarily have to use 10 and 1 \solum\ as isotropic luminosities to discriminate between the kilomaser and megamaser regime for \water\ and OH, respectively. According to this classification, in all the galaxies where OH maser emission has been found,  95 \% (89/94) are OH megamasers. As far as water masers are concerned, 73 \% (62/85) are megamasers. Fig.~\ref{fig:rel} indicates that, for the subsample of galaxies searched for both maser species, most of the OH detections (14 out of 19) are megamasers (with NGC\,1068 and NGC\,4355 having luminosities very close to the upper limit for kilomasers). Since very luminous masers are always found in AGN, the distances for the OH maser galaxy subsample is, on average, higher than that of the \water\ masers, where a relatively lower number of megamaser sources (21 out of 39) is present. Due to the different average distance of the samples, the luminosity detection threshold is typically higher for water maser searches in OH-masing galaxies than that for OH searches in \water-maser hosts. This may partly justify the results for quadrants I to III: no OH megamaser with a detected \water\ kilomaser but 4 OH kilomasers, indeed the closest four galaxies of the OH subsample, detected also in the water maser line. This suggests that weak masers from both species should indeed coexist in the same object (although possibly not at the very same location), similarly to what happens in galactic star forming regions. Possibly, only a selection effect due to the sensitivity of the surveys has, so far, limited the number of detections of the contemporary presence of kilomasers from both species to a few. 

In our study, only two galaxies show megamaser emission from both molecular species, Arp\,299 and UGC\,5101. However, the OH megamaser detection in UGC\,5101 reported by \citet{martin89} was never confirmed, and doubts were raised by the non-detection obtained later by \citet{baan92}. Without a confirmation of the OH megamaser line in UGC\,5101, we consider the system Arp\,299 as the only clear case with detected megamaser emission from both molecular species. But then, what is causing this apparent lack of 'double' detections? To answer this question a more systematic approach (parallel surveys to detect \water\  maser emission in OH maser galaxies, and viceversa) and a more detailed study of the, so far, unique case, IC\,694, are necessary.

%______________________________________________________________

\section{Conclusions}\label{concl}

From the results of our interferometric observations on three extragalactic water maser galaxies and by more extended studies of water (and OH) maser sources, we can draw the following main conclusions:
\begin{itemize}
\item{The kilomaser in NGC\,3556 has a star formation origin associated with a compact radio continuum source, possibly identified as a supernova remnant or radio supernova.}
\item{The megamaser in Arp\,299 originates from three distinct regions associated with the nuclear regions of the two merging galaxies, NGC\,3690 and IC\,694, and from their region of interaction.}
\item{Our study on Arp\,299 confirms the likely presence of an AGN in the nucleus of NGC\,3690. Without ruling out the possibility for AGN activity in IC\,694, our outcome favours instead an association for the maser in this galaxy with an expanding structure possibly related to the extreme star formation activity taking place in the innermost region of the galaxy.}
\item{The kilomaser in NGC\,4151 originates in the nuclear region. Its association with the AGN, known to be present in the galaxy, and hence, its nature, cannot, however, be confidently assessed because we only tentatively detect one of the two maser features in our maps. The known variability of the two line features is possibly the main cause for our partial detection.}
\item{Kilomasers are mostly associated with massive star formation, although a number of them is credibly related to low-luminosity AGN activity. The nuclear X-ray column density for the former type is typically lower than that for the AGN-associated candidates, strengthening the possibility of the existence of two distinct classes. Given, however, the low-luminosity of some of the maser sources associated with AGN, our study reinforces the conviction that a distinction between SF- and AGN-associated water masers based on a sharp isotropic luminosity threshold is not proper and requires instead case studies of individual systems to be assessed.} 
\item{A thorough analysis of the interplay between extragalactic maser emission from \water\ and OH is affected by the low number of objects searched for both transitions. Nevertheless, our study indicates that, as expected, weaker maser emission from both molecular species can indeed be present in the same object, most likely because of the association of such emission with star formation processes. Only one case, that of Arp\,299, is found to host megamaser emission from both molecules. This case deserves indeed to be studied in even more detail.}
\end{itemize}
%
%______________________________________________________________

\begin{acknowledgements}
We wish to thank the referee for his/her comments on the manuscript. GS is supported by the Deutsche Forschungsgemeinschaft (DFG) through the Emmy Noether Reasearch grant VL 61/3-1.
This research has made use of the NASA/IPAC Extragalactic Database (NED), which is operated by the Jet Propulsion Laboratory, Caltech, under contract with NASA. This research has also made use of the NASA's Astrophysics Data System Abstract Service (ADS).
\end{acknowledgements}

\bibliographystyle{aa} % style aa.bst
\bibliography{followups_ph} % your references (.bib)

%\clearpage

\onllongtab{5}{
%\longtab{1}{
\begin{longtable}{llcccccc}
 \caption{\label{waterandoh} All galaxies known so far to host either \water (standard font) or OH (italics) maser emission, or both (bold face). The lack of information on the maser luminosity (OH or \water) means that, to the best of our knowledge, neither a detection nor an upper limit for that transition was reported in the literature.}\\
\hline\hline
  IRAS              & Alt. & RA     & Dec.  & Distance      & $log(\rm {L{^{iso}_{\rm {H_{2}O}}}})$ & $log(\rm {L{^{iso}_{\rm OH}}})$ & Ref.$^{a}$ \\
  name              & name  &        &         & (Mpc)    &     (\solum)               &  (\solum)            &      \\
\hline
\endfirsthead
\caption{continued.}\\
\hline\hline
   IRAS              & Alt. & RA     & Dec.        & Distance & $log(\rm {L{^{iso}_{\rm {H_{2}O}}}})$ & $log(\rm {L{^{iso}_{\rm OH}}})$ &  Ref.$^{a}$  \\
   name              & name &        &             &  (Mpc)   &  (\solum)               &  (\solum)            &       \\
\hline
\endhead
\hline
\endfoot
  {\it {00057+4021}}  &                     &    00:08:20.5 &  +40:37:57  &13389  &      $-$        &  1.93   &  $-$,h10   \\
  00073+2538   & NGC23              &    00:09:53.6 &  +25:55:23  &     4566    &  2.2           &  $<$-1.0          & w2,h11  \\
%%  00085--1223  & 2 NGC17            &    00:11:06.5 &  -12:06:26  &     5881    &  $-$           &  $<$0.88          & unpub,h1\&h3  \\
%%  N/A          & 3 J0011-0054       &    00:11:45.2 &  -00:54:31  &    14384    &  $-$           &                     & unpub,  \\
  00175+5902   & IC10	            &    00:20:17.3 &  +59:18:14  &     -348      &  -0.74         &  $<$-3.6         & w1,h2  \\
%%  {\it {00335--2732}} &                     &    00:36:00.5 &  -27:15:34  &20771  &                &  $-$    &  ,h10   \\ 
  N/A          & NGC235A          &    00:42:52.8 &  -23:32:28  &     6664      &  2.0           &     $-$                & w3,$-$  \\
{\bf {00450--2533}}  & {\bf {NGC253}}           &    00:47:33.1 &  -25:17:18    &      243   &  -0.74         &  -2.0            & w1,h6  \\
  00461+3141   & Mrk348           &    00:48:47.1 &  +31:57:25  &     4507      &  2.6           &  $<$-0.67         & w1,h11  \\
%%  00510--0901  & 8 NGC291           &    00:53:29.9 &  -08:46:04  &     5705    &  $-$           &                     & unpub,  \\
%%  {\it {00509+1225}}  &   {\it {UGC545}}  &    00:53:34.9 &  +12:41:36  &18330  & &  $-$    &  ,h10   \\ 
  01063--8034  & ESO013-G012      &    01:07:02.2 &  -80:18:28  &     5047   &  2.7           &      $-$               & w1,$-$  \\
  01133+3249   & Mrk1            &    01:16:07.2 &  +33:05:22  &     4780   &  1.7           &      $-$               & w1,$-$  \\
  01219+0331   & NGC520          &    01:24:35.1 &  +03:47:33  &     2281   &  0.0           &  $<$-1.5          & w5,h11  \\ 
%%  N/A          & 12 J0126-0417 	    &    01:26:01.7 &  -04:17:56  &     5639   &  $-$           &                     & unpub,  \\
  01310+3024   & M33             &    01:33:50.9 &  +30:39:36  &     -179   &  -0.46         &  $<$-4.8         & w1,h4  \\
  01306+3524   & NGC591          &    01:33:31.2 &  +35:40:06  &     4547   &  1.4           &     $-$                & w1,$-$  \\
  01319--2940  & NGC613          &    01:34:18.2 &  -29:25:07  &     1481   &  1.2--1.5      &    $-$                 & w3--w5,$-$  \\
  {\it {01364--1042}} &                     &    01:38:52.9 &  -10:27:11  &14464  &  $<$ 0.56        &  1.60   &  w7,h10   \\   
  {\it {01418+1651}}  &    {\it {III Zw35}}         &    01:44:30.5 &  +17:06:05  & 8225  &  $<$ 1.38             &  2.39   &  w3,h10   \\ 
  {\it {01562+2527}}  &                     &    01:59:02.6 &  +25:42:37  &49665  &      $-$          &  2.99   & $-$ ,h10   \\  
  01573--0704  & IC0184          &    01:59:51.2 &  -06:50:25  &     5382   &  1.4           &       $-$              & w3,$-$  \\
%%  N/A          & 17 J0214-0016      &    02:14:05.9 &  -00:16:37  &    11205   &  $-$           &                     & unpub,  \\   
  02386--0828  & NGC1052         &    02:41:04.8 &  -08:15:21  &     1510   &  2.1           &  $<$ 0.31         & w1,h1  \\
{\bf {02401--0013}}  & {\bf {NGC1068}}         &    02:42:40.7 &  -00:00:48  &     1137   &  2.2           &       -0.08        & w1,h5  \\
  02474+4127   & NGC1106         &    02:50:40.5 &  +41:40:17  &     4337   &  0.90          &       $-$              & w2,$-$  \\
%%  {\it {02483+4302}}  &                     &    02:51:35.8 &  +43:15:12  &15421  &                &  $-$    &  ,h10   \\ 
%%  N/A          & 21 J0253-0014      &    02:53:29.6 &  -00:14:06  &     8622   &  $-$           &                     & unpub,  \\  
  {\it {02524+2046}}  &                     &    02:55:17.1 &  +20:58:43  &54383  &     $-$           &  3.35   & $-$ ,h10   \\  
  02568+3637   & Mrk1066         &    02:59:58.6 &  +36:49:14  &     3605   &  1.5           &  $<$0.20         & w1,h3  \\
%%  03012--0117  & 23 NGC1194         &    03:03:49.1 &  -01:06:13  &     4076   &  $-$           &                     & unpub,  \\  
%%  {\it {03056+2034}}  &    {\it {UGC2553}}          &    03:08:30.7 &  +20:46:20  &8225   &                &  $-$    &  ,h10   \\ 
%%  03222--0313  & 24 NGC1320         &    03:24:48.7 &  -03:02:32  &     2663   &  $-$           &                     & unpub,  \\  
  {\it {03260--1422}} &                     &    03:28:24.3 &  -14:12:07  &12736  &        $-$        &  2.04   & $-$ ,h10   \\ 
%%  N/A          & 25 J0336-0750 	    &    03:36:46.2 &  -07:50:24  &    11719   &  $-$           &                     & unpub,  \\
  03348--3609  & NGC1386         &    03:36:46.4 &  -36:00:02  &     868    &  2.1           &      $-$               & w1, $-$ \\
  03355+0104   & IRAS03355+0104  &    03:38:10.4 &  +01:14:18  &   11926    &  2.6           &      $-$               & w7,$-$  \\
  03419+6756   & IC342           &    03:46:48.5 &  +68:05:46  &       31   &  -2            &  $<$-2.9          & w1,h3  \\
  {\it {03521+0028}}  &                     &    03:54:42.2 &  +00:37:03  & 45541 &        $-$        &  1.95   & $-$ ,h10   \\ 
  {\it {03566+1647}}  &                     &    03:59:29.1 &  +16:56:26  & 40029 &        $-$        &  1.61   & $-$ ,h10   \\  
  N/A          & J0414+0534 	    &    04:14:37.8 &  +05:34:42  &    z=2.64  &  4             &     $-$               & w6, $-$ \\
  {\it {04121+0223}}  &                     &    04:14:47.1 &  +02:30:36  &36702  &         $-$       &  2.29   &  ,h10   \\  
  {\it {04332+0209}}  &    {\it {UGC3097}}          &    04:35:48.4 &  +02:15:29  &3590   &     $-$           &  0.53   & $-$ ,h10   \\ 
%%  04385--0828  & 30 IRAS04385-0828  &    04:40:51.0 &  -08:22:22  &   4527     &  $-$           &                     & unpub,  \\
  {\it {04454--4838}} & {\it {ESO 203-IG-001}}      &    04:46:49.5 &  -48:33:33  &15862  &      $-$          &  2.51   & $-$ ,h10   \\  
  04502+0258   & UGC3193         &    04:52:52.7 &  +03:03:24  &   4454     &  2.4           &     $-$                & w2, $-$ \\
%%  N/A          & 32CGCG468-002Ned01 &    05:08:19.7 &  +17:21:48  &   5248     &  $-$           &                     & unpub,  \\
  05071+0725   & UGC3255         &    05:09:50.2 &  +07:29:00  &   5669     &  1.2           &      $-$               & w1, $-$ \\ 
  {\it {05100--2425}} &                     &    05:12:09.2 &  -24:21:56  &10047  &    $-$            &  1.35   & $-$ ,h10   \\  
%%  {\it {05414+5840}}  &   {\it {UGC3351}}           &    05:45:47.9 &  +58:42:04  &4455   &   $<$ 0.98            &  $-$    &  w3,h10   \\   
  06097+7103   & Mrk3            &    06:15:36.3 &  +71:02:15  &     4050   &  1.0           &  $<$0.47         & w1,h3  \\
  06106+7822   & NGC2146         &    06:18:37.7 &  +78:21:25  &      893   &  0.3           &  $<$-1.5         & w1,h2  \\
  {\it {06206--3646}} &                     &    06:22:22.4 &  -36:47:43  &32390  &    $-$            &  2.32   & $-$ ,h10   \\  
  06256+6342   & VIIZw073        &    06:30:25.6 &  +63:40:41  &  12391     &  2.2           &       $-$              & w3, $-$ \\
  06456+6054   & NGC2273         &    06:50:08.7 &  +60:50:45  &     1840   &  0.8           &  $<$-0.82         & w4,h2  \\
  {\it {06487+2208}}  &                     &    06:51:45.8 &  +22:04:27  &42972  &       $-$         &  2.90   & $-$ ,h10   \\   
  {\it {07163+0817}}  &                     &    07:19:05.5 &  +08:12:07  &33269  &      $-$          &  1.47   & $-$ ,h10   \\ 
  07151+5926   & UGC3789         &    07:19:30.9 &  +59:21:18  &   3325     &  2.6           &      $-$               & w2, $-$ \\ 
%%  07318+3255   & 39 NGC2410         &    07:35:02.2 &  +32:49:20  &   4681     &  $-$           &                     & unpub,  \\  
  07379+6517   & Mrk78           &    07:42:41.7 &  +65:10:37  &  11137     &  1.5           &       $-$              & w1, $-$ \\
  {\it {07572+0533}}  &                     &    07:59:57.2 &  +05:25:00  &56900  &      $-$          &  2.63   & $-$ ,h10   \\ 
%%  07572+2650   & 41 IC0485          &    08:00:19.8 &  +26:42:05  &   8338     &  $-$           &                     & unpub,  \\  
  08014+0515   & Mrk1210         &    08:04:05.8 &  +05:06:50  &   4046     &  1.9           &          $-$           & w1,$-$  \\ 
  N/A          & SDSSJ0804+3607  &    08:04:31.0 &  +36:07:18  &    z=0.66  &  4.36          &          $-$           & w8,$-$  \\
  {\it {08071+0509}}  &                     &    08:09:47.2 &  +05:01:09  &15650  &         $-$       &  1.90   & $-$ ,h10   \\ 
  {\it {08201+2801}}  &                     &    08:23:12.6 &  +27:51:40  &50314  &          $-$      &  3.30   &$-$  ,h10   \\ 
  {\it {08279+0956}}  &                     &    08:30:40.9 &  +09:46:28  &62547  &          $-$      &  2.98   &$-$  ,h10   \\  
  08341--2614  & He 2-10         &    08:36:15.2 &  -26:24:34  &   873      &  0.68          &     $-$                & w9, $-$ \\
  N/A          & 2MASXJ0836+3327 &    08:36:22.8 &  +33:27:39  &  14810     &  3.5           &      $-$               & w7,$-$  \\
  08400+5023   & NGC2639         &    08:43:38.1 &  +50:12:20  &     3336   &  1.4           &  $<$-0.50         & w1,h2  \\ 
  {\it {08449+2332}}  &                     &    08:47:50.2 &  +23:21:10  &45406  &       $-$         &  2.12   & $-$ ,h10   \\  
  {\it {08474+1813}}  &                     &    08:50:18.3 &  +18:02:01  &43591  &       $-$         &  2.66   & $-$ ,h10   \\ 
  {\it {09039+0503}}  &                     &    09:06:34.2 &  +04:51:25  &37516  &       $-$         &  2.63   & $-$ ,h10   \\ 
%%  N/A          & 47 J0912+2304 	    &    09:12:46.4 &  +23:04:27  &  10889     &  $-$           &                     & unpub,  \\
%%  09091--1436  & 48 NGC2781 	    &    09:11:27.5 &  -14:49:01  &  2053      &  $-$           &                     & unpub,  \\
  09108+4019   & NGC2782         &    09:14:05.1 &  +40:06:49  &     2543   &  1.1           &  $<$-0.65         & w1,h2  \\ 
  09160+2628   & NGC2824         &    09:19:02.2 &  +26:16:12  &   2760     &  2.7           &    $-$                 & w1, $-$ \\  
  09277+4917   & SBS0927+493     &    09:31:06.7 &  +49:04:47  &  10167     &  2.4           &   $-$                  & w7,$-$  \\
  {\bf {09320+6134}}   & {\bf {UGC5101 (?; see text)}}         &    09 35 51 6 &  +61 21 11  &  11802     &  3.2           &   1.73            & w4,h10  \\
%%  {\it {09320+6134}}  &  {\it {UGC5101}} &    09:35:51.6 &  +61:21:11  &11802  &   detezione ($<$ 1.35) &  1.73   &  w12,h10   \\  
  09380+0348   & Mrk1419         &    09:40:36.4 &  +03:34:37  &  4932      &  2.6           &       $-$              & w1,$-$  \\
  09407--1009  & NGC2979         &    09:43:08.6 &  -10:23:00  &  2720      &  2.1           &      $-$               & w1, $-$ \\
  09430--1808  & NGC2989         &    09:45:25.8 &  -18:22:36  &  4146      &  1.6           &       $-$              & w2,$-$  \\
  {\it {09531+1430}}  &                     &    09:55:51.1 &  +14:16:01  &64538  &     $-$           &  3.01   & $-$ ,h10   \\  
{\bf {09517+6954}}   & {\bf {NGC3034}}-{\bf {M82}}  &    09:55:52.7 &  +69:40:46  &   203      &  0.0           &        -2.1       & w1,h6  \\
  {\it {09539+0857}}  &                     &    09:56:34.3 &  +08:43:06  &38643  &        $-$        &  3.26   &$-$  ,h10   \\ 
%%  N/A          & 57 NGC3081         &    09:59:29.5 &  -22:49:35  &  2391      &  $-$           &                     & unpub,  \\ 
  {\bf {09585+5555}}   & {\bf {NGC3079}}         &    10:01:57.8 &  +55:40:47  &     1116   &  2.7           &   -1.1            & w1,h8  \\
  {\it {10039--3338}} &{\it {IC2545}}-{\it {ESO374-IG-032}} &    10:06:04.8 &  -33:53:15  &10223  &  $<$ 1.88             &  2.86   &  w14,h10   \\  
  {\it {10036+2740}}  &                     &    10:06:26.3 &  +27:25:46  &49466  &        $-$        &  1.99   &$-$  ,h10   \\   
%%  N/A          & 59 J1011-1926 	    &    10:11:50.6 &  -19:26:44  &    8058    &  $-$           &                     & unpub,  \\
%%  10109+3905   & 60 NGC3160         &    10:13:55.1 &  +38:50:34  &   6920     &  $-$           &                     & unpub,  \\ 
  10140--3318  & IC2560          &    10:16:18.7 &  -33:33:50  &   2925     &  2.0           &            $-$         & w1, $-$ \\
  {\it {10173+0828}}  &                     &    10:20:00.2 &  +08:13:34  &14716  &      $-$          &  2.46   & $-$ ,h10   \\ 
  10257--4338  & NGC3256         &    10:27:51.3 &  -43:54:14  &     2804   &  1.04          &  $<$0.65          & w10,h7  \\
%%  10288+2614   & 63 UGC5713 	    &    10:31:38.9 &  +25:59:02  &   6312     &  $-$           &                     & unpub,  \\
  10309+6017   & Mrk34           &    10:34:08.6 &  +60:01:52  &  15140     &  3.0           &           $-$          & w1,$-$  \\
  {\it {10339+1548}}  &                     &    10:36:37.9 &  +15:32:42  &59130  &         $-$       &  2.34   & $-$ ,h10   \\ 
  {\it {10378+1109}}  &                     &    10:40:29.2 &  +10:53:18  &40854  &         $-$       &  3.20   & $-$ ,h10   \\  
  10433+6329   & NGC3359         &    10:46:36.8 &  +63:13:25  &     1014   &  -0.15         &  $<$-1.5         & w2,h2  \\  
  10459--2453  & NGC3393         &    10:48:23.4 &  -25:09:43  &   3750     &  2.4--2.6      &         $-$            & w3--w4, $-$ \\
%%  {\it {10485--1447}} &                     &    10:51:03.1 &  -15:03:22  &39872  &                &  $-$    &  ,h10   \\  
%%  N/A          & 67 UGC6093         &    11:00:48.0 &  +10:43:41  & 10828      &  $-$           &                     & unpub,  \\
  {\it {11010+4107}}  &    {\it {Arp148}}           &    11:03:53.2 &  +40:50:57  & 10350 &    $-$            &  1.74   &$-$  ,h10   \\  
  {\it {11029+3130}}  &                     &    11:05:37.5 &  +31:14:32  &59540  &      $-$          &  2.53   &$-$  ,h10   \\ 
%%  N/A          & 68 J1109+2837 	    &    11:09:33.1 &  +28:37:39  &  11422     &  $-$           &                     & unpub,  \\
  11085+5556   & NGC3556         &    11:11:31.0 &  +55:40:27  &      699   &  0.0           &  $<$-1.7         & w1,h2  \\
  11143--7556  & NGC3620         &    11:16:04.7 &  -76:12:59  &     1680   &  0.32          & $<$0.04           & w10,h7  \\
%%   N/A         & 71 CGCG185-028     &    11:17:00.1 &  +32:35:51  &  10455     &  $-$           &                     & unpub,  \\
%%  11155+4601   & 72 NGC3614 	    &    11:18:21.3 &  +45:44:54  &  2333      &  $-$           &                     & unpub,  \\
  {\it {11180+1623}}  &                     &    11:20:41.7 &  +16:06:57  &49766  &       $-$         &  2.17   &$-$  ,h10   \\  
 {\bf { 11257+5850}}   & {\bf {Arp299}}          &    11:28:31.9 &  +58:33:45  &     3088   &  2.1           & 1.1                 & w1,h3  \\
  11330+7048   & NGC3735         &    11:35:57.3 &  +70:32:09  &     2696   &  1.1           & $<$-0.50          & w17,h2  \\
%%  11355+1128   & 75 IRASF11355+1128 &    11:38:08.0 &  +11:11:47  &  10707     &  $-$           &                     & unpub,  \\ 
%%  11365--3727  & 76 NGC3783         &    11:39:01.7 &  -37:44:19  &  2917      &  $-$           &                     & unpub,  \\
%%  11470+5048   & 77 MCG+09-19-205   &    11:49:45.7 &  +50:31:37  &  7924      &  $-$           &                     & unpub,  \\ 
  {\it {11506--3851}} & {\it {ESO320-G-030}}        &    11:53:11.7 &  -39:07:49  &3232   &   $<$1.66             &  1.52   &  w15,h10   \\  
  11593--1835  & Antennae        &    12:01:53.3 &  -18:52:37  &     1705   &  8.2           & $<$-0.51          & w9,h2  \\
  {\it {11524+1058}}  &                     &    11:55:02.8 &  +10:41:44  &53564  &       $-$         &  2.81   &$-$  ,h10   \\  
%%  N/A          & 79 J1202+3519 	    &    12:02:04.6 &  +35:19:18  & 10201      &  $-$           &                     & unpub,  \\
%%  11598+1507   & 80 UGC7016         &    12:02:24.0 &  +14:50:37  &  7271      &  $-$           &                     & unpub,  \\ 
  {\it {12005+0009}}  &                     &    12:03:04.4 &  -00:07:27  &36646  &      $-$          &  2.02   & $-$ ,h10   \\   
  12005+4448   & NGC4051         &    12:03:09.6 &  +44:31:53  &      700   &  0.3           & $<$-1.5         & w1,h6  \\
  {\it {12018+1941}}  &                     &    12:04:24.5 &  +19:25:10  &50559  &    $-$            &  2.49   &$-$  ,h10   \\ 
  {\it {12032+1707}}  &                     &    12:05:47.7 &  +16:51:08  &65291  &     $-$           &  4.15   &$-$  ,h10   \\ 
  12080+3940   & NGC4151         &    12:10:32.6 &  +39:24:21  &      995   &  -0.2          & $<$-1.4          & w1,h2  \\
  {\it {12112+0305}}  &                     &    12:13:46.0 &  +02:48:38  &21980  &  $<$ 1.89             &  2.78   &  w13,h10   \\  
  12131+3636   & NGC4214         &    12:15:39.2 &  +36:19:37  &      291   &  0.028         &  $<$-2.6         & w9,h2  \\
%%  12159+3005   & 84 NGC4253 	    &    12:18:26.5 &  +29:48:46  &  3876      &  $-$           &                     & unpub,  \\
  {\it {12162+1047}}  &                     &    12:18:47.7 &  +10:31:11  &43920  &       $-$         &  2.05   & $-$ ,h10   \\  
  N/A          & NGC4258         &    12:18:57.5 &  +47:18:14  &      448   &  1.9           & $<$-2.1          & w1,h2  \\
  12186+1839   & NGC4293         &    12:21:12.9 &  +18:22:57  &      893   &  0.1--0.7      & $<$-2.1          & w3,h4  \\
  12232+1256   & NGC4388         &    12:25:46.7 &  +12:39:44  &     2524   &  1.1           & $<$-1.3         & w1,h11  \\
  {\it {12243--0036}} &     {\it {NGC4355}}         &    12:26:54.6 &  -00:52:39  &2179   &  $<$ -0.14      &  -0.01  &  w5,h10   \\ 
  12315+0255   & NGC4527         &    12:34:08.5 &  +02:39:14  &   1736     &  0.60          & $-$                 & w2, $-$ \\
  {\it {12540+5708}}  &{\it {Mrk231 - UGC8058}}     &    12:56:14.2 &  +56:52:25  &12642  &  $<$ 1.18             &  2.84   &  w4,h10   \\  
  N/A          & ESO269-G012     &    12:56:40.5 &  -46:55:34  &  5014      &  3.0           &     $-$                & w1,$-$  \\
  {\it {12548+2403}}  &                     &    12:57:20.0 &  +23:47:46  &39483  &     $-$           &  1.89   & $-$ ,h10   \\  
  12590+2934   & NGC4922         &    13:01:25.2 &  +29:18:50  &     7071   &  2.3           & $<$-0.47           & w1,h11  \\
  13025--4911  & NGC4945         &    13:05:27.5 &  -49:28:06  &      563   &  1.7           & $<$-1.5             & w1,surcis  \\
%%  13044--2324  & 92 NGC4968 	    &    13:07:06.0 &  -23:40:37  &  2957      &  $-$           &                     & unpub,  \\
%%  {\it {13097--1531}} &       {\it {NGC5010}}       &    13:12:26.3 &  -15:47:52  &2975   &                &  $-$    &  ,h10   \\ 
  {\it {13218+0552}}  &                     &    13:24:19.9 &  +05:37:05  &61488  &           $-$     &  3.07   & $-$ ,h10   \\ 
%%  {\it {13254+4754}}  &                     &    13:27:29.7 &  +47:39:22  &18107  &                &  $-$    &  ,h10   \\       
  13277+4727   & NGC5194 (M51)   &    13:29:52.7 &  +47:11:43  &      463   &  -0.2          & $<$-2.0          & w1,h2  \\  
  13362+4831   & NGC5256         &    13:38:17.2 &  +48:16:32  &      8353  &  1.5           & $<$0.27           & w1,h2  \\
  13370--3123  & NGC5253         &    13:39:55.9 &  -31:38:24  &   407      &  0.021         &     $-$                & w9,$-$  \\
  {\it {13428+5608}}  &{\it {Mrk273}}-{\it {UGC8696}}     &    13:44:42.1 &  +55:53:13  &11326  &   $<$ 0.37             &  2.55   &  w16,h10   \\ 
%%  N/A          & 96 SBS1344+527     &    13:46:40.8 &  +52:28:37  &  8763      &  $-$           &                     & unpub,  \\  
  {\it {13451+1232}}  &   {\it {4C +12.50}}         &    13:47:33.3 &  +12:17:24  &36497  &   $<$ 2.25            &  2.28   &  w13,h10   \\   
  13510+3344   & NGC5347         &    13:53:17.8 &  +33:29:27  &  2335      &  1.5           &      $-$               & w1, $-$ \\
%%  N/A          & 98 J1355+0553 	    &    13:55:35.9 &  +05:53:05  &  11762     &  $-$           &                     & unpub,  \\
%%  13550+6452   & 99 MCG+11-17-010   &    13:56:28.7 &  +64:37:43  &   9456     &  $-$           &                     & unpub,  \\ 
  {\it {14043+0624}}  &                     &    14:06:49.8 &  +06:10:36  &33933  &         $-$       &  1.68   &$-$  ,h10   \\  
  {\it {14059+2000}}  &                     &    14:08:18.7 &  +19:46:23  &37084  &         $-$       &  2.97   &$-$  ,h10   \\ 
%%  14057--2920  & 100 ESO446-G018    &    14:08:38.3 &  -29:34:19  &   4771     &  $-$           &                     & unpub,  \\
  {\it {14070+0525}}  &                     &    14:09:31.2 &  +05:11:31  &79259  &         $-$       &  3.88   &$-$  ,h10   \\  
  14095--2652  & NGC5495        &    14:12:23.3 &  -27:06:29  &   6737     &  2.3           &        $-$             & w3,$-$  \\
  14092--6506  & Circinus       &    14:13:09.3 &  -65:20:21  &    434     &  1.3           &        $-$             & w1,$-$  \\
  14106--0258  & NGC5506        &    14:13:14.8 &  -03:12:27  &     1853   &  1.7           & $<$-0.47          & w1,h2  \\
  14294--4357  & NGC5643        &    14:32:40.8 &  -44:10:29  &     1199   &  1.4           &      $-$               & w1,$-$  \\
  14396--1702  & NGC5728        &    14:42:23.9 &  -17:15:11  &     2804   &  1.9           & $<$-0.33          & w1,h2  \\
  14547+2448   & UGC9618B       &    14:57:00.7 &  +24:37:03  &   10094    &  2.8           &          $-$           & w7,$-$  \\
  {\it {14553+1245}}  &                     &    14:57:43.4 &  +12:33:16  &37444  &    $-$            &  1.94   & $-$ ,h10   \\  
%%  14568+4504   & 107 UGC9639        &    14:58:35.5 &  +44:53:06  &  10802     &  $-$           &                     & unpub,  \\  
  14566--1629  & NGC5793        &    14:59:24.7 &  -16:41:36  &  3491      &  2.0           &         $-$            & w1, $-$ \\
  {\it {14586+1431}}  &                     &    15:01:00.4 &  +14:20:15  &44279  &        $-$        &  2.54   &$-$  ,h10   \\  
  {\it {15065--1107}} &       {\it {NGC5861}}       &    15:09:16.1 &  -11:19:18  &1851   &      $-$          &  0.04   & $-$ ,h10   \\ 
  {\it {15107+0724}}  &  {\it {CGCG049-057}}        &    15:13:13.1 &  +07:13:32  &3897   &   $<$ 0.43            &  0.85   &  w13,h10   \\  
  {\it {15179+3956}}  &                     &    15:19:47.1 &  +39:45:38  &14261  &        $-$        &  1.02   & $-$ ,h10   \\   
%%  N/A          & 109 2MASXJ15201+52 &    15:20:19.6 &  +52:53:56  & 11166      &  $-$           &                     & unpub,  \\  
  {\it {15224+1033}}  &                     &    15:24:51.5 &  +10:22:45  &40187  &         $-$       &  2.22   & $-$ ,h10   \\  
  {\it {15247--0945}} &                     &    15:27:27.8 &  -09:55:41  &11993  &         $-$       &  1.59   & $-$ ,h10   \\   
  {\it {15250+3609}}  &                     &    15:26:59.4 &  +35:58:38  &16535  &         $-$       &  2.52   & $-$ ,h10   \\  
  {\it {15327+2340}}  &      {\it {Arp220}}         &    15:34:57.1 &  +23:30:11  &5434   &  $<$ -0.27             &  2.54   &  w16,h10   \\ 
  {\it {15587+1609}}  &                     &    16:01:03.6 &  +16:01:03  &41126  &        $-$        &  3.04   &$-$  ,h10   \\ 
%%  N/A          & 110 2MASXJ16070+01 &    16:07:03.9 &  +01:06:29  &  8216      &  $-$           &                     & unpub,  \\  
  {\it {16100+2528}}  &                     &    16:12:05.4 &  +25:20:23  &39688  &        $-$        &  1.68   &$-$  ,h10   \\   
%%  {\it {16145+4231}}  &                     &    16:16:11.6 &  +42:23:59  &6944   &                &  $-$    &  ,h10   \\           
  {\it {16255+2801}}  &                     &    16:27:38.1 &  +27:54:52  &40056  &      $-$          &  2.38   & $-$ ,h10   \\   
%%  16288+3929   & 111 IRAS16288+3929 &    16:30:32.6 &  +39:23:03  & 9161       &  $-$           &                     & unpub,  \\  
%%  16287+3035   & 112 CGCG168-018    &    16:30:40.9 &  +30:29:20  &11015       &  $-$           &                     & unpub,  \\
  {\it {16300+1558}}  &                     &    16:32:21.4 &  +15:51:46  &72467  &       $-$         &  2.71   &$-$  ,h10   \\  
  {\it {16399--0937}} &                     &    16:42:40.2 &  -09:43:14  &8098   &       $-$         &  1.68   &$-$  ,h10   \\    
  16504+0228   & NGC6240        &    16:52:58.9 &  +02:24:03  &     7339   &  1.6           & $<$-0.50           & w1,h11  \\
  16552+2755   & NGC6264        &    16:57:16.1 &  +27:50:59  & 10177      &  2.5           &            $-$         & w7,$-$  \\
%%  N/A          & 115 J1658+3923     &    16:58:15.5 &  +39:23:29  & 10290      &  $-$           &                     & unpub,  \\ 
  17118+4350   & NGC6323        &    17:13:18.0 &  +43:46:56  & 7772       &  2.7           &         $-$            & w1,$-$  \\
  17123--6245  & NGC6300        &    17:16:59.5 &  -62:49:14  & 1109       &  0.5           &          $-$           & w1,$-$  \\
  {\it {17160+2006}}  &                     &    17:18:15.6 &  +20:02:58  &32917  &           $-$     &  2.04   &$-$  ,h10   \\   
  {\it {17208--0014}} &                     &    17:23:21.9 &  -00:17:01  &12834  &            $-$    &  3.04   &$-$  ,h10   \\ 
  {\it {17526+3253}}  &     {\it {UGC11035}}        &    17:54:29.4 &  +32:53:14  &7798   &   $-$             &  0.81   &$-$  ,h10   \\   
  {\it {17540+2935}}  &                     &    17:55:56.1 &  +29:35:26  &32410  &        $-$        &  1.56   & $-$ ,h10   \\  
  18333--6528  & ESO103-G35     &    18:38:20.3 &  -65:25:42  & 3983       &  2.6           &            $-$         & w1,$-$  \\
  {\it {18368+3549}}  &                     &    18:38:35.4 &  +35:52:20  &34827  &           $-$     &  2.81   & $-$ ,h10   \\  
  {\it {18544--3718}} &                     &    18:57:52.7 &  -37:14:38  &22012  &           $-$     &  2.23   & $-$ ,h10   \\   
  {\it {18588+3517}}  &                     &    19:00:41.2 &  +35:21:27  &31996  &           $-$     &  2.12   & $-$ ,h10   \\   
  19370--0131  & IRASF19370-0131 &    19:39:38.9 &  -01:24:33  & 6622       &  2.2           &             $-$        & w1, $-$ \\
  19497+0222   & 3C403          &    19:52:15.8 &  +02:30:24  & 17688      &  3.3           &             $-$        & w1, $-$ \\
  {\it {20100--4156}} &                     &    20:13:29.5 &  -41:47:35  &38848  &   $<$ 4.44            &  3.77   &  w15,h10   \\  
  {\it {20248+1734}}  &                     &    20:27:08.9 &  +17:44:20  &36513  &         $-$       &  2.23   & $-$ ,h10   \\  
  {\it {20286+1846}}  &                     &    20:30:55.5 &  +18:56:46  &40700  &          $-$      &  3.19   &$-$  ,h10   \\ 
  20305--0211  & NGC6926        &    20:33:06.1 &  -02:01:39  & 5880       &  2.7           &            $-$         & w1, $-$ \\
%  20338+5958   & NGC6946        &    20:34:52.3 &  +60:09:14  &       48   &  0.0           & $<$-2.4          & w11,h2  \\
  {\it {20450+2140}}  &                     &    20:47:14.3 &  +21:51:12  &38489  &        $-$        &  2.12   & $-$ ,h10   \\  
%%  {\it {20491+1846}}  &     {\it {UGC11643}}        &    20:51:25.9 &  +18:58:04  &8735   &                &  $-$    &  ,h10   \\          
  {\it {20550+1656}}  &      {\it {II Zw96}}        &    20:57:23.9 &  +17:07:39  &10822  &    $-$            &  1.92   &$-$  ,h10   \\   
  {\it {21077+3358}}  &                     &    21:09:49.0 &  +34:10:20  &52973  &           $-$     &  2.94   & $-$ ,h10   \\  
  {\it {21272+2514}}  &                     &    21:29:29.4 &  +25:27:50  &45208  &          $-$      &  3.36   & $-$ ,h10   \\  
  21583--3800  & AM2158-380b    &    22:01:17.1 &  -37:46:24  & 9983       &  2.7           &    $-$                 & w3,$-$  \\
  {\it {22025+4205}}  &      {\it {UGC11898}}       &    22:04:36.1 &  +42:19:38  &4290   &    $-$            &  0.86   &$-$  ,h10   \\  
  {\it {22055+3024}}  &                     &    22:07:49.7 &  +30:39:40  &38041  &          $-$      &  2.36   & $-$ ,h10   \\   
  {\it {22088--1832}} &                     &    22:11:33.8 &  -18:17:06  &51026  &          $-$      &  2.59   & $-$ ,h10   \\   
  {\it {22116+0437}}  &                     &    22:14:09.9 &  +04:52:24  &58098  &          $-$      &  2.26   & $-$ ,h10   \\  
  22265--1826  & IRASF22265-1826 &    22:29:12.5 &  -18:10:47  & 7520       &  3.8           &         $-$            & w1, $-$ \\
  {\it {22491--1808}} &                     &    22:51:49.2 &  -17:52:23  &23312  &       $-$         &  1.82   & $-$ ,h10   \\   
  {\it {23019+3405}}  &                     &    23:04:21.2 &  +34:21:48  &32389  &       $-$         &  1.78   & $-$ ,h10   \\   
  23024+1203   & NGC7479        &    23:04:56.7 &  +12:19:22  &      2381  &  1.3           & $<$-1.3         & w2,h6  \\
  {\it {23028+0725}}  &                     &    23:05:20.4 &  +07:41:44  &44784  &          $-$      &  3.00   & $-$ ,h10   \\  
  {\it {23129+2548}}  &                     &    23:15:21.4 &  +26:04:32  &53637  &          $-$      &  3.10   &$-$  ,h10   \\  
  {\it {23135+2516}}  &     {\it {IC5298}}          &    23:16:00.7 &  +25:33:24  &8221   &  $<$ 1.11             &  0.80   &  w3,h10   \\  
  23168+0537   & IC1481         &    23:19:25.1 &  +05:54:22  &  6118      &  2.5           &             $-$        & w1, $-$  \\
  {\it {23199+0122}}  &                     &    23:22:31.6 &  +01:39:28  &40679  &    $-$            &  2.05   &$-$  ,h10   \\   
  {\it {23233+0946}}  &                     &    23:25:56.2 &  +10:02:49  &38356  &      $-$          &  2.55   & $-$ ,h10   \\  
  {\it {23365+3604}}  &                     &    23:39:01.3 &  +36:21:09  &19331  &     $-$           &  2.30   & $-$ ,h10   \\  

\hline
%\end{tabular}
%\end{minipage}
%}}
\end{longtable}
$^{a}$ References: (w1) Henkel et al., 2005, A\&A, 436, 75; (w2) Braatz \& Gugliucci, 2008, ApJ, 678, 96; (w3) Kondratko et al., 2006, ApJ, 638, 100; (w4) Zhang et al., 2006, A\&A, 450, 933; (w5) Castangia et al., 2008, A\&A, 479, 111; (w6) Impellizzeri et al., 2008, Nature, 456, 927; (w7) Kondratko et al., 2006, ApJ, 652, 136 (luminosities are derived directly from the spectra); (w8) Barvainis \& Antonucci, 2005, ApJ, 628, 89; (w9) Darling et al., 2008, ApJ, 685, 39; (w10) Surcis et al., 2009, A\&A, 502, 529; (w11) Claussen et al., 1984, Nature, 310, 298; (w12) Hagiwara et al., 2002, A\&A, 383, 65; (w13) Hagiwara et al., 2003, MNRAS, 344, 53; (w14) Greenhill et al., 2002, ApJ, 565, 836; (w15) Braatz et al., 1996, ApJS, 106, 51; (w16) Braatz et al., 2004, ApJ, 617, 29; (w17) Greenhill et al., 1997, ApJ, 486, 15; (h1) Norris et al., 1989, MNRAS, 237, 673; (h2) Baan et al., 1992, AJ, 103, 728; (h3) Staveley-Smith et al., 1987, MNRAS, 226, 689; (h4) Schmelz \& Baan, 1988, AJ, 95, 672; (h5) Gallimore et al., 1996, ApJ, 462, 740; (h6) Unger et al., 1986, MNRAS, 220, 1; (h7) Staveley-Smith et al., 1992, MNRAS, 258, 725; (h8) Baan \& Irwin, 1995, ApJ, 446, 602; (h9) Baan et al., 1982, ApJ, 260, 49; (h10) Kl\"{o}ckner, 2004, Ph.D.\ Thesis, and references therein; (h11) Schmelz et al., 1986, AJ, 92, 1291}

\end{document}